\documentclass[twocolumn,superscriptaddress,floatfix,preprintnumbers,hyperref,nofootinbib,nobibnotes,amsmath,amssymb,aps]{revtex4-2}

\usepackage{graphicx}
\usepackage{subcaption}
\usepackage{dcolumn}
\usepackage{bm}
\usepackage{hyperref}
\usepackage[mathlines]{lineno}
\usepackage{color}
\usepackage{xcolor}
\usepackage{verbatim}
\usepackage{soul}

\definecolor{brown}{rgb}{.7,.35,.1}

\begin{document}

\preprint{FERMILAB-PUB-25-0053-T}

\title{High-Frequency Gravitational Waves on BREAD}

\author{Rodolfo Capdevilla}
\email{rcapdevi@fnal.gov}
\affiliation{Theoretical Physics Division, Fermi National Accelerator Laboratory, Batavia, IL 60510, USA}
\author{Roni Harnik}
\email{roni@fnal.gov}
\affiliation{Theoretical Physics Division, Fermi National Accelerator Laboratory, Batavia, IL 60510, USA}
\affiliation{Superconducting Quantum Materials and Systems Center (SQMS), Fermilab, Batavia, IL 60510, USA}
\author{Taegyun Kim}
\email{tkim12@alumni.nd.edu}
\affiliation{Uichang District Office, 468, Dogye-dong Uichang-gu, Changwon, 51381, South Korea}
\affiliation{Department of Physics, Southern University of Science and Technology, Shenzhen, 518055, China}
\author{Tom Krokotsch}
\email{tom.krokotsch@desy.de}
\affiliation{Universität Hamburg, Luruper Chaussee 149, 22761 Hamburg, Germany}

\date{\today}

\begin{abstract}
The use of experiments searching for light axion dark matter as high-frequency gravitational wave (HFGW) detectors has garnered increasing attention in recent years. We explore the capabilities of the Broadband Reflector Experiment for Axion Detection (BREAD) in probing the GW parameter space and study the directional dependence of its coverage. This detector can investigate frequencies ranging from 0.05 to 200 THz. We find that employing single photon detectors BREAD is sensitive to GWs with characteristic strains as low as \(10^{-21}\) at 0.1 THz and~\(10^{-25}\) at 200 THz with a year exposure time, making it competitive with other proposals operating at similar frequencies.
\end{abstract}

\maketitle

\section{Introduction}
\label{sec:intro}


The discovery of gravitational waves (GWs) has opened a new channel in the multi-messenger exploration of the Universe. The interferometers of LIGO and Virgo have demonstrated that gravitational perturbations propagate through spacetime in the form of waves~\cite{LIGOScientific:2016aoc}. These perturbations arise from large, time-varying quadrupole moments taking place during very violent events, such as the merger of black holes and neutron stars. The typical frequencies of these events fall between $10^{-6}$ to $10^4$ Hz. In analogy with electromagnetic radiation, it might be possible that gravitational radiation also exists across a broad range of frequencies throughout the Universe. Like with light, detecting this radiation across a wide frequency spectrum can teach us a great deal. For this reason, designing new experimental concepts and proposing theoretically motivated sources for high-frequency GWs (HFGWs) is an active field of research~\cite{Aggarwal:2020olq,Aggarwal:2025noe}.


A variety of potential sources of GW with frequencies higher than 1 MHz can be identified through mechanisms involving physics within and beyond the Standard Model (SM). Within the SM one can identify sources like stars~\cite{McDonald:2024nxj,Garcia-Cely:2024ujr} and the primordial plasma of the Universe~\cite{Ghiglieri:2015nfa,Giovannini:2019oii,Ghiglieri:2020mhm,Ringwald:2020ist,Giovannini:2023itq}. It is well known that plasmas can produce GWs through microscopic graviton emission and macroscopic hydrodynamic fluctuations~\cite{Cabral:2016klm,Garg:2022wdm,Garg:2023yaw}. In the case of neutron stars photon-graviton conversion can happen resonantly at the magnetosphere~\cite{McDonald:2024nxj}. The primordial plasma produced a spectrum of GW with frequencies equal and below $\sim 100$ GHz  and with a characteristic strains below $h_0\sim10^{-34}$, though this prediction depend on assumptions about the early Universe like the reheating temperature~\cite{Ringwald:2020ist}. When it comes to physics beyond the SM (BSM) one can find a variety of potential sources of HFGW. One example is light primordial black holes (PBHs) ~\cite{Dolgov:2011cq,Franciolini:2022htd}, which can either merge or evaporate producing a background of GW with frequencies up to GHz. Dark sector physics can feature either exotic compact objects~\cite{Giudice:2016zpa} or particles that decay into gravitons~\cite{Landini:2025jgj}, where the latter can produce a spectrum of optical gravitons in the cosmos. Primordial magnetic fields~\cite{Fujita:2020rdx} can convert thermalized photons from the early Universe into a sizable population of gravitons. Other mechanisms include cosmic strings~\cite{Servant:2023tua}, modified gravity~\cite{Delgado:2025ext}, Inflation~\cite{Bernal:2023wus,Garcia:2024zir,Saha:2024lil}, and black hole superradiance~\cite{Arvanitaki:2009fg,Arvanitaki:2010sy,Yoshino:2013ofa}.


To detect HFGWs, one can employ experimental setups that are highly sensitive to mechanical effects, including optical sensors~\cite{Arvanitaki:2012cn,Aggarwal:2020umq,Carney:2024zzk}, phonons~\cite{Goryachev:2014yra,Kahn:2023mrj}, cavities~\cite{Ballantini:2005am, Berlin:2023grv}, magnets~\cite{Domcke:2024mfu}, and, of course, interferometers~\cite{Schnabel:2024hem}. All these detectors rely on the deformations caused by GWs in the experimental setup. An alternative approach is to focus on the electromagnetic effects produced by GWs. For instance, GWs can couple to matter and induce spin transitions~\cite{Quach:2016uxd} or excite plasmons~\cite{Ito:2019wcb}. Another example involves experiments where GWs are converted into electromagnetic signals through the Gertsenshtein effect~\cite{Gertsenshtein:1961xxx}. This is the gravitational equivalent of the Sikivie effect
, where axions convert into photons in the presence of a background magnetic field~\cite{Sikivie:1983ip}. In this context, recent years have seen significant exploration of HFGW sensitivity using axion dark matter (DM) experiments across a wide range of frequencies: from MHz to GHz~\cite{Domcke:2022rgu,Bringmann:2023gba,Domcke:2023bat}, between GHz and THz~\cite{Berlin:2021txa,Navarro:2023eii,Ahn:2023mrg,Gatti:2024mde,Domcke:2024eti,Capdevilla:2024cby}, and beyond THz~\cite{Ejlli:2019bqj}. The motivation for studying the potential detection of GWs in axion DM experiments is two-fold: First, in many cases, these searches only require a straightforward recasting of current or foreseeable data. Second, since current sensitivity in the GW parameter space primarily covers regions far from known SM sources, discovering a GW signal above GHz with sufficiently high strain would most likely constitute a BSM discovery.


In this paper, we explore the sensitivity of the Broadband Reflector Experiment for Axion Detection (BREAD) to HFGWs. This experiment can probe a wide range of frequencies above 10 GHz, possibly up to 300 THz, depending on the detector technology used~\cite{BREAD:2021tpx}. BREAD belongs to the category of broadband dish antennas designed for searches for light new physics~\cite{Horns:2012jf,Suzuki:2015sza,Jaeckel:2015kea,Knirck:2018ojz,FUNKExperiment:2020ofv,Tomita:2020usq}. The setup consists of a cylindrical metal barrel with a fully reflective internal surface and a coaxial parabolic reflector that focuses signal photons onto a sensor positioned at the top of the cylinder. The entire system is immersed in a strong magnetic field, with its geometry specifically designed for optimal placement in standard cryostats and compact high-field solenoids. 

The BREAD experiment was designed to detect the conversion of axion dark matter to photons, which like GWs, can occur in a magnetic field. It is worthwhile noting a key difference between the two processes. The axion is a massive particle, and so its conversion to photons requires violation of energy or momentum conservation. In the case of BREAD, the breaking of translation invariance by the reflective cylinder walls absorbs the required momentum to phase-match axion-to-photon conversion. The effective volume for the conversion process therefore scales with the area of the reflector, times the wavelength of the light (set by the axion mass),~$V_\mathrm{eff}\sim L^2\lambda$. By contrast, the gravitational wave is massless and is kinematically allowed to convert to a photon, so long as the background magnetic field breaks rotational symmetry, enabling spin-2 to spin-1 conversion. The effective volume for conversion is thus of order the detector volume,~$V_\mathrm{eff}\sim L^3$.

Our task will be to compute the sensitivity projections on the parameter space of GWs in the case of a constant emitter in the cosmos, or a background of GWs from the early Universe.
This paper is organized as follows: in section~\ref{sec:calculations} we discuss the experimental setup and compute the rate for GW conversion into photons inside BREAD under our simplifying assumptions. In section~\ref{sec:focusing} we discuss the focusing efficiency of BREAD in the case of GWs. In section~\ref{sec:results} we show our projected sensitivity to monochromatic and stochastic GWs in the wide range of frequencies accessible to BREAD. In section \ref{sec:conclusion}, we conclude by discussing our simplifying assumptions and potential ideas for enhancing the sensitivity of BREAD to GWs.

\begin{figure}[t]
\centering
\includegraphics[width=0.4\textwidth]{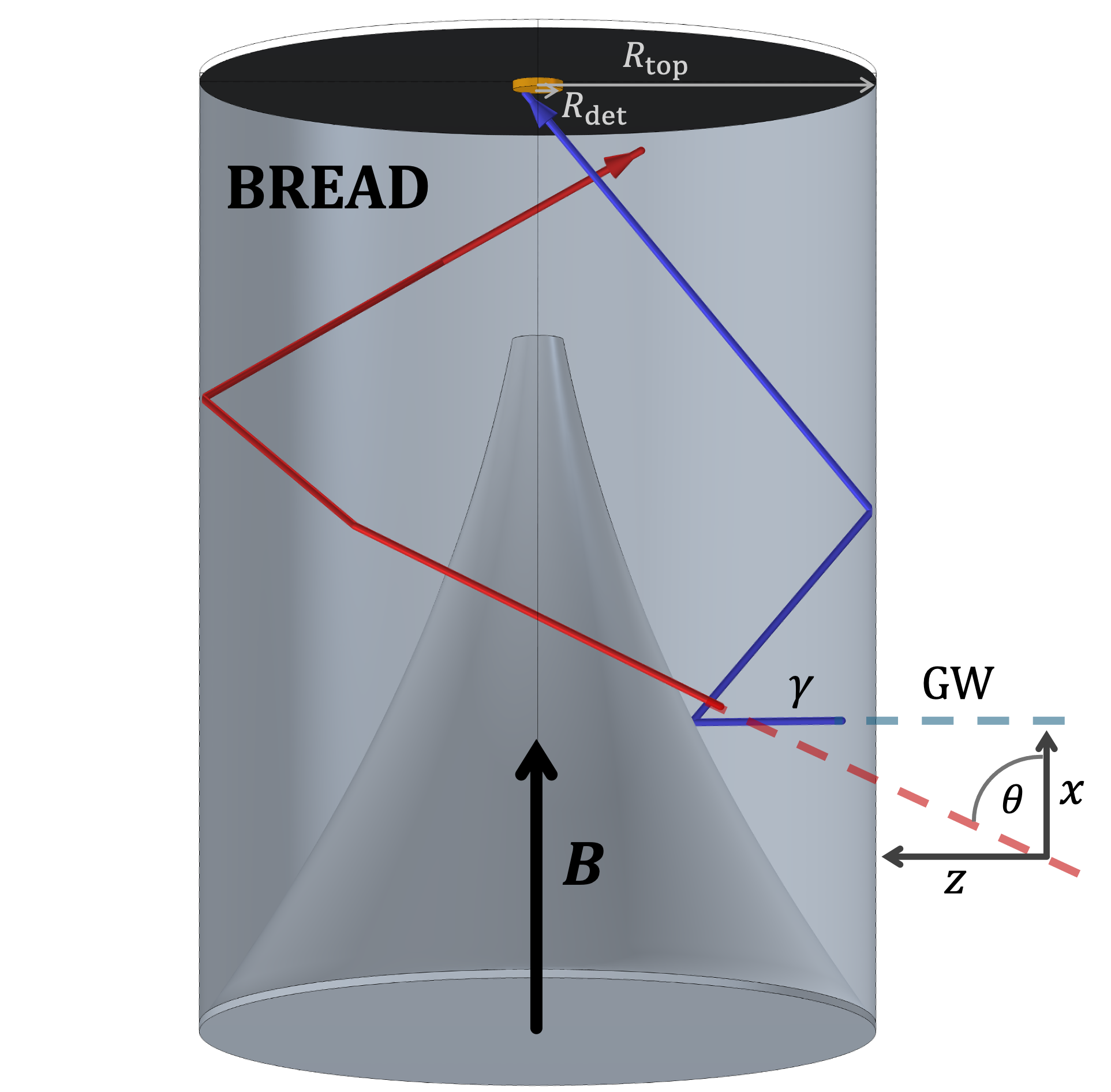}
\caption{Schematic description of our setup. The GW (dashed) rays travel along an angle $\theta$ with the $x$-axis to the symmetry axis of the cylinder which coincides with the external $\bm{B}$-field direction. As the GW overlaps with the magnetic field, signal photons emerge within the full detector volume (this is in contrast to signal photons for axion dark matter which emerge near the surface). The blue ray illustrates a case where the photon is focused to the detector area in orange. The red ray shows a case where the signal does not get focused.}
\label{fig:cartoon}
\end{figure}

\section{Calculations}
\label{sec:calculations}


The physical process we aim to describe is illustrated in Fig.~\ref{fig:cartoon}. A GW originates from a distant source emitting with angular frequency $\omega_g$. Starting with a simple example first, we assume that the GW fronts reach BREAD, with the cavity axis oriented perpendicular to the direction of the incoming wave (blue rays in Fig.~\ref{fig:cartoon}). The magnetic field ($ B$) is aligned along the cavity axis. When the GW interacts with the $ B$ field, a signal photon is produced and directed to the detector by the focusing system of BREAD. The signal photon can, in principle, emerge at any angle relative to the GW direction. However, at very high frequencies, the photon predominantly follows the same direction as the incident GW. This can be understood through conservation of energy and momentum during the GW-to-photon conversion, which limits the transverse momentum imparted to the photon to be (at most) $1/L$, where $L$ is the characteristic length of the experimental setup. In our regime of interest $\omega_gL \gg 1$, the GW$\to\gamma$ conversion occurs highly collinearly, where the photon inherits the direction of the incoming GW.


We begin our calculation by considering the free-photon action with explicit dependence on the metric $g^{\mu\nu}$, which we expand as $\eta^{\mu\nu}-h^{\mu\nu}$, where $\eta$ is the flat metric diag$(-+++)$ and $h$ is a small perturbation. With this we get
\begin{eqnarray}
\nonumber
 S &=& \int d^4x\sqrt{-g}\left[-\frac{1}{4}g^{\mu\nu}g^{\alpha\beta}F_{\mu\alpha}F_{\nu\beta}\right], \\
 &\supset& \int d^4x\,\,\, h^{\lambda\sigma}P_{\mu\nu\lambda\alpha\sigma\beta}F^{\mu\alpha}\bar{F}^{\nu\beta},\label{eq:action}
\end{eqnarray}
where $\bar{F}^{\nu\beta}$ describes the background $ B$ field and $F^{\mu\alpha}$ the signal photon. The tensor $P$ is given by
\begin{equation}
P_{\mu\nu\lambda\alpha\sigma\beta} = \eta_{\mu\nu}\eta_{\lambda\alpha}\eta_{\sigma\beta} - \frac{1}{4} \eta_{\lambda\sigma}\eta_{\mu\nu}\eta_{\alpha\beta}.
\end{equation}

\begin{figure}[t]
\centering
\includegraphics[width=0.25\textwidth]{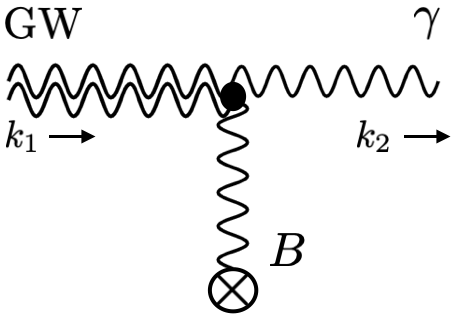}
\caption{Scattering of an incoming GW with momentum $k_1$ off a background $B$ field producing an outgoing photon with momentum $k_2$.}
\label{fig:diagram}
\end{figure} 

Our goal now is to compute the transition probability in Fig.~\ref{fig:diagram}. Effectively, the process at hand can be seen as an incoming GW with momentum $k_1=\omega_1(1,\hat {\bf k}_1)\equiv \omega_1\hat k_1$ scattering off a $ B$ field in a box of length $L$ producing a photon with momentum $k_2=\omega_2(1,\hat {\bf k}_2)\equiv \omega_2\hat k_2$. The amplitude of the transition can be written as
\begin{equation}
\mathcal{A}_{fi}=\left\langle \gamma_{k_2} \right| i \int d^4x\, h^{\lambda\sigma}P_{\mu\nu\lambda\alpha\sigma\beta}F^{\mu\alpha}\bar{F}^{\nu\beta} \left| 0 \right\rangle
\label{eq:amplitude}
\end{equation}
where the final state is the outgoing photon and there are no on-shell
photons in the initial state. Instead, the initial state vacuum is permeated by the overlap of the incoming GW and the background $ B$ field. In this calculation we treat both as classical fields, whereas the outgoing photon is quantized within a volume $V_\gamma$
\begin{equation}
\left\langle 0\right|  A_{\mu}(x)\left|\gamma_{k_{2}}\right\rangle =\frac{1}{\sqrt{2V_{\gamma}\omega_2}}\epsilon_{\mu}(k_{2})~e^{ik_{2}\cdot x},
\label{eq:photon}
\end{equation}
where $\epsilon_{\mu}(k_{2})$ is the polarization vector of the photon.


In order to compute the space integral in Eq.~\eqref{eq:amplitude} we need an explicit form for the tensors $h$ and $\bar F$. The latter is trivial, as it is just the electromagnetic tensor describing a $ B$ field in the $x$-axis
\begin{equation}
\bar{F}^{\nu\beta}=\left(\begin{array}{cccc}
0 & 0 & 0 & 0\\
0 & 0 & 0 & 0\\
0 & 0 & 0 & 1\\
0 & 0 & -1 & 0
\end{array}\right)B({\bf x})\equiv \hat F^{\nu\beta}B({\bf x}).
\end{equation}
To describe the spatial dependence of the GW, we need to choose a specific gauge. In this work, we describe the tensor $h$ in both: the transverse-traceless (TT) gauge, where the system falls freely under the gravitational pull of the GW, and the proper detector (PD) frame~\cite{Manasse:1963zz}, which employs Fermi normal coordinates with the origin fixed at the center of mass of the detector. The choice between TT gauge and PD frame draws questions as each choice is more suitable for different ranges of the  product of the GW frequency and the characteristic length of the system. It has been argued in recent literature that the TT frame is well-suited for the limit of high frequencies $\omega_g L \gg 1$ where the different components of the system react as free-falling objects and the background B field is considered static~\cite{Ratzinger:2024spd,Domcke:2024eti}. In~\cite{Ratzinger:2024spd} it is argued that the PD frame is more suitable in the limit $\omega_g L \ll 1$, where the assumption of rigidity of the system is well justified. Below we perform the calculation in both frames and show that in the high frequency limit the two rates agree.

\subsection{Transverse-Traceless Gauge}


We perform our calculation in the limit $\omega_g L \gg 1$. Under this condition, the incoming GW perceives the system as an ``infinite'' box, allowing boundary effects to be neglected in principle. In the TT gauge, an incoming GW traveling along the $z$-axis can be expressed as
\begin{eqnarray}
   h^{\mu\nu}_{\rm TT}(x) &=& \frac{1}{\sqrt{2}} \left(\begin{array}{cccc}
0 & 0 & 0 & 0\\
0 & h_{+} & h_{\times} & 0\\
0 & h_{\times} & -h_{+} & 0\\
0 & 0 & 0 & 0
\end{array}\right)
e^{i k_1\centerdot x}, \nonumber \\
&\equiv&\frac{h_0}{\sqrt{2}} \hat h^{\mu\nu}_{p} e^{i k_1\centerdot x},
\label{eq:hTT}
\end{eqnarray}
where $h_0$ represents the strain or amplitude of the GW and the label $p$ indicates either the $+$ and $\times$ polarizations. For example, the matrix $\hat h^{\mu\nu}_{+}=$ diag$(0,1,-1,0)$.


Under these considerations the amplitude in Eq.~\eqref{eq:amplitude} can be written as
\begin{equation}
\mathcal{A}_{fi}= \frac{h_0}{2} \sqrt{\frac{\omega_2}{V_\gamma}} T_{pp'} \hat B({\bf q})(2\pi)\delta(\omega_2-\omega_1),
\label{eq:amplitude_2}
\end{equation}
where $p'$ is the photon polarization, ${\bf q} = {\bf k}_2-{\bf k}_1$ and
\begin{eqnarray}
T_{pp'}&=&\hat h_p^{\lambda\sigma}P_{\mu\nu\lambda\alpha\sigma\beta}(\hat k_2^\mu\epsilon_{p'}^{*\alpha}-\hat k_2^\alpha\epsilon_{p'}^{*\mu})\hat F^{\nu\beta}, \label{eq:TppTT} \\
\hat B({\bf q})&=&\int d^3x B({\bf x})e^{-i({\bf k}_2-{\bf k}_1)\centerdot{\bf x}}.\label{eq:B}
\end{eqnarray}
To compute the transition probability $W$ we need to integrate over the phase space of the final state photon
\begin{equation}
W = \int \frac{d^3{\bf k}_2}{(2\pi)^3} V_\gamma |\mathcal{A}_{fi}|^2.
\label{eq:prob}
\end{equation}
In the {\it large box limit} that we are interested in, Eq.~\eqref{eq:B} simplifies to $\hat B({\bf q})=B(2\pi)^3\delta({\bf q})$. As we integrate over ${\bf k}_2$ this delta function forces ${\bf q}\to0$ which in turn implies $\omega_2\to\omega_1$ so that $\int d^3x e^{-i{\bf q}\centerdot{\bf x}} \to V$ and $\int dt e^{i(\omega_2-\omega_1)t}\to T$, where $V$ is the volume enclosing the $B$ field and $T$ is the transit time that a GW wavefront takes to pass through the system. Inserting these formulas, along with Eq.~\eqref{eq:amplitude_2} into Eq.~\eqref{eq:prob} we can write the transition probability of a GW with polarization $p$ into a photon of polarization $p'$ as
\begin{equation}
W_{pp'} = \frac{h_0^2B^2}{4} \omega_g |T_{pp'}|^2 V T^2.
\label{eq:prob_2}
\end{equation}
The rate is related to this probability simply by $R=W/T$. Now, to get the total rate we sum over the final state polarizations of the photon while averaging over the initial polarizations of the GW. Because of our large box limit the photon is fully collinear with the GW, which travels in the $z$ direction. Therefore, the polarization states of the photon have a very simple form $\epsilon_\pm = (0,\mp1,-i,0)/\sqrt{2}$, and with this, the total rate is given by
\begin{eqnarray}
R_{\rm tot}^\text{max}&=& \frac{1}{2} \sum_{pp'} R_{pp'},\nonumber \\
           &=& \frac{1}{4} \omega_g h_0^2B^2VT,   \label{eq:Rtot}
\end{eqnarray}
which agrees with the classic result in \cite{Gertsenshtein:1961xxx} and more explicitly with the rate found in \cite{PhysRevD.16.2915}.

In terms of power,  Eq.~(\ref{eq:Rtot}) gives a signal power of $P_{\rm GW} = \omega R = \omega^2 h_0^2 B^2 V L/4.$ 
It is interesting to compare this to the power deposited in BREAD by axion dark matter as given in \cite{BREAD:2021tpx}, 
\begin{equation}
P_{\rm axion} = \frac{\rho_{\rm DM} B^2 g_{a\gamma\gamma}^2 A}{2 m_a^2}
= \frac{m_a^2 (a g_{a\gamma\gamma})^2 B^2 A L^2}{4 (m_a L)^2}\,,
\end{equation}
where we have used the relation of the axion field amplitude \( a \) to the dark matter energy density $\rho_\mathrm{DM}$. Using a map \( h_0 \leftrightarrow a g_{a\gamma\gamma} \) and $m_a\leftrightarrow\omega$, we find that $P_{\rm axion} = P_{\rm GW}/(\omega L)^2.$ This reflects, as we discussed above, that GWs can covert to photons in the full volume of BREAD, whereas axion conversions only occurs near the boundary, and hence scales with the area $A$, accounting for one power of $\omega L$. A second power of $\omega L$ comes from the scaling of the GW transit time with $L$.
These comparisons highlight that, although BREAD?s focusing system is not optimized for signal photons produced isotropically, as in the case for the GW signal, the fact that GW-to-photon conversion occurs throughout the entire effective volume leads to a sizable signal. In essence, axion DM searches at BREAD benefit from the focusing system, while GW searches benefit from the large conversion volume, which makes BREAD a competitive experiment for both types of searches.

\subsection{Proper Detector Frame}


Following~\cite{Berlin:2021txa}, we describe the metric perturbation in the PD frame using their resumed expression valid to all orders in ${\bf k}\cdot {\bf x}$, given by

\begin{align}
    h^{\text{PD}}_{00}&= \omega_0^2 \, F({\bf k} \cdot {\bf x}) \,{\bf b} \cdot {\bf x} \\
    h^{\text{PD}}_{0i}&= \frac{\omega_0^2}{2} \left[F({\bf k} \cdot {\bf x})- i F'({\bf k} \cdot {\bf x})  \right] \left( \hat{\bf k} \cdot {\bf x} \, b_i - {\bf b} \cdot {\bf x} \,\hat{k}_i \right)\\
    h^{\text{PD}}_{ij}&= \frac{\omega_0^2}{i}F'({\bf k} \cdot {\bf x}) \Big( |{\bf x}|^2 h^\text{TT}_{ij}|_{{\bf x} = 0} + {\bf b} \cdot {\bf x} \,\delta_{ij}\nonumber \\
    &  ~~~~~~~~~~~~~~~~~~~~~~~~~~~~~~~~~ - b_i x_j - b_jx_i \Big),
\end{align}
where
\begin{align}
    b_j &= x_i \left( h^{\text{TT}}_{ij}|_{{\bf x} = 0}  \right) \\
    F(\xi) &= \frac{e^{i \xi}-1-i\xi}{\xi^2},
\end{align}
the vector $x_i$ describes the coordinates of the experimental setup with origin in its center of mass and $F'$ is the derivative of $F$ with respect to its argument.

In analogy to Eq.~\eqref{eq:hTT} we describe the tensor $h$ in terms of multiple factors
\begin{equation}
h_\text{PD}^{\mu\nu} \equiv \frac{h_0}{\sqrt{2}}\, \hat{h}^{\mu\nu}({\bf x})\, e^{-i\omega_g t}.    
\end{equation}
With this, the amplitude can be written as
\begin{equation}
\mathcal{A}_{fi}^\text{PD} = \frac{\pi h_0B\omega_2}{\sqrt{V_\gamma\omega_2}} \delta(\omega_2-\omega_1)\int d^3x~T_{pp'}(z)~e^{-i {\bf k}_2\cdot {\bf x}}.
\label{eq:amplitude_PD}
\end{equation}
Just like in the previous section, in the limit of high frequency, we compute the rate assuming collinearity between the photon and the GW. We evaluate the tensor product $T_{pp'}$ using Eq.~\eqref{eq:TppTT} which in PD frame displays spatial dependence as follows:
\begin{align}
    T_{+\pm}(z) =& \frac{2+iz\omega_g+e^{iz\omega_g}(-2+iz\omega_g)}{\sqrt{2}\,z\,\omega_g}\label{eq:TppmPD}\\
    T_{\times\pm}(z)=&\pm\frac{-2i+z\omega_g+e^{iz\omega_g}(2i+z\omega_g)}{\sqrt{2}\,z\,\omega_g}.\label{eq:TcpmPD}
\end{align}
With these expressions we now compute the rate in the PD frame. Because $T_{pp'}$ does not depend on the transverse coordinates, the amplitude is proportional to delta functions that force the conservation of the transverse momentum. These delta function in turn lead to length factors in the rate $\int dr e^{ir(k_{2,r}-k_{1,r})}\to L_r$, where $r=x,y$ are the transverse coordinates. The rate is now given by
\begin{align}
    R_{pp'}^\text{PD}&=\frac{h_0^2 B^2}{4}\omega_g L_x\,L_y\,\left| \int dz e^{-i\omega_g z} ~ T_{pp'}(z)\right|^2\\
    &=\frac{h_0^2 B^2}{8}\omega_g L_x\,L_y\,L_z^2\left(1+ \frac{1}{L_z\omega_z}f(L_z\omega_g) \right)^2\\
    &=R_{pp'}^\text{TT} \left(1+ \frac{1}{L_z\omega_z}f(L_z\omega_g) \right)^2\label{eq:RppPD}
\end{align}
where
\begin{align}
    f(\zeta) &= 2\sin\left( \zeta/2\right) -4\,\text{Si}\left( \zeta/2 \right),\nonumber \\
    \text{Si}(\zeta) &=\int_0^\zeta \frac{\sin \zeta'}{\zeta'}d\zeta'.\nonumber
\end{align}
Note that in the limit $\zeta = L_z\omega_g\gg1$, we have that $\text{Si}(\zeta)\to\pi/2$, and $R_{pp'}^\text{PD}\to R_{pp'}^\text{TT}$. As argued in \cite{Ratzinger:2024spd}, the overlap of the GW and the $B$ field in the PD frame should include the effect of the detector motion as the experimental setup is not at rest in said frame in the $\omega_g L\gg1$ limit. For example, one should include the small radiative component of the background magnetic field due to the mechanical oscillation of the coils that generate said field~\cite{Domcke:2024mfu}, as well as the effects of deformations of the walls of the cylinder as mechanical resonances could be excited. However, the radiative component of the $B$ field from the coil deformations is shielded by the walls of the cavity (as also argued in~\cite{Berlin:2021txa}). Note that the mechanical motion of the cavity walls can be a dominant effect in setups where the cavity is loaded with microwave power~\cite{Ballantini:2005am,Berlin:2023grv}. In our setup, BREAD, both effects are negligible, and hence the calculations agree in both frames.

\section{Focusing}
\label{sec:focusing}

In the previous section we computed the rate for a GW entering BREAD to convert to a photon.
However, any GW detector utilizing photon regeneration in magnetic fields, not only needs to create a large amount of signal photons by maximizing Eq.~\eqref{eq:Rtot}, but also find a way to read out as many of them as possible. Dish antennas like BREAD achieve this by using reflectors that focus axion induced photons onto a small focal point. However, unlike in axion electrodynamics, gravitationally induced photons with $\omega_gR\gg1$ are emitted in the same direction as the incoming GW and not necessarily perpendicular to the cylindrical walls. Since the focusing mechanism depends on this, the signal-to-detection efficiency $\epsilon_s$ is heavily affected. 

To investigate the focusing of GW induced photons in a BREAD geometry, we use the Geometrical Optics Module in COMSOL Multiphysics\textsuperscript{\tiny\textregistered} \cite{comsol} to perform ray tracing simulations, which is a valid approximation in the $\sim\text{THz}$ frequency range. The actual focal area through which the detector can measure power flux depends on the specific detection scheme employed. For example, the GigaBREAD detector \cite{Knirck:2018ojz, GigaBREAD_Axion_search} at radio frequencies (RF) $\sim 10\,\text{GHz}$ uses a coaxial horn antenna. Since we are mostly considering higher frequencies, we leave the detector method general and define the focal area as the fraction of the top of the cylinder for which the parabolic mirror is flattened $R_\text{det}^2/R_\text{top}^2\simeq0.0056$ (see Fig.~\ref{fig:cartoon}). For the simulation, we release $N_\text{tot}=10^5$ rays, evenly distributed in the volume at an angle $\theta$ to the symmetry axis of the cylinder, where $\theta=0$ corresponds to $\hat{\bm{k}}_2\parallel \hat{\bm{x}}$. Afterwards, we count the amount of rays $N_\text{focus}$ hitting the focal area and estimate the signal-to-detection efficiency $\epsilon_s(\theta)=N_\text{focus}(\theta)/N_\text{tot}$. Since there is no value of $\theta$ at which GWs are expected a priori, we need to consider a sky-averaged value for our sensitivity estimate.

For a GW arriving at an angle $\theta$ with respect to the symmetry axis of the cylinder, the rate in Eq.~\eqref{eq:Rtot} gets modified to
\begin{equation}
R_{\rm tot}=\frac{1}{4}\omega_gh_0^2B^2VT\sin^2\theta.
\end{equation}
We now define our sky-averaged signal-to-detection efficiency by
\begin{equation}
\bar{\epsilon}_s=\frac{\langle\epsilon(\theta)R_\text{tot}(\theta)\rangle_\theta}{R_\text{tot}(\pi/2)},    
\end{equation}
which takes the entire angular dependence of the detector into account so that the sky-averaged rate of signal photons reaching the focal area becomes
\begin{equation}
\langle R_\text{sig}\rangle=\bar{\epsilon}_sR_\text{tot}(\pi/2).    
\end{equation}

\begin{figure}[t]
    \centering
    \begin{subfigure}{0.4\textwidth}
        \centering
        \includegraphics[width=\textwidth]{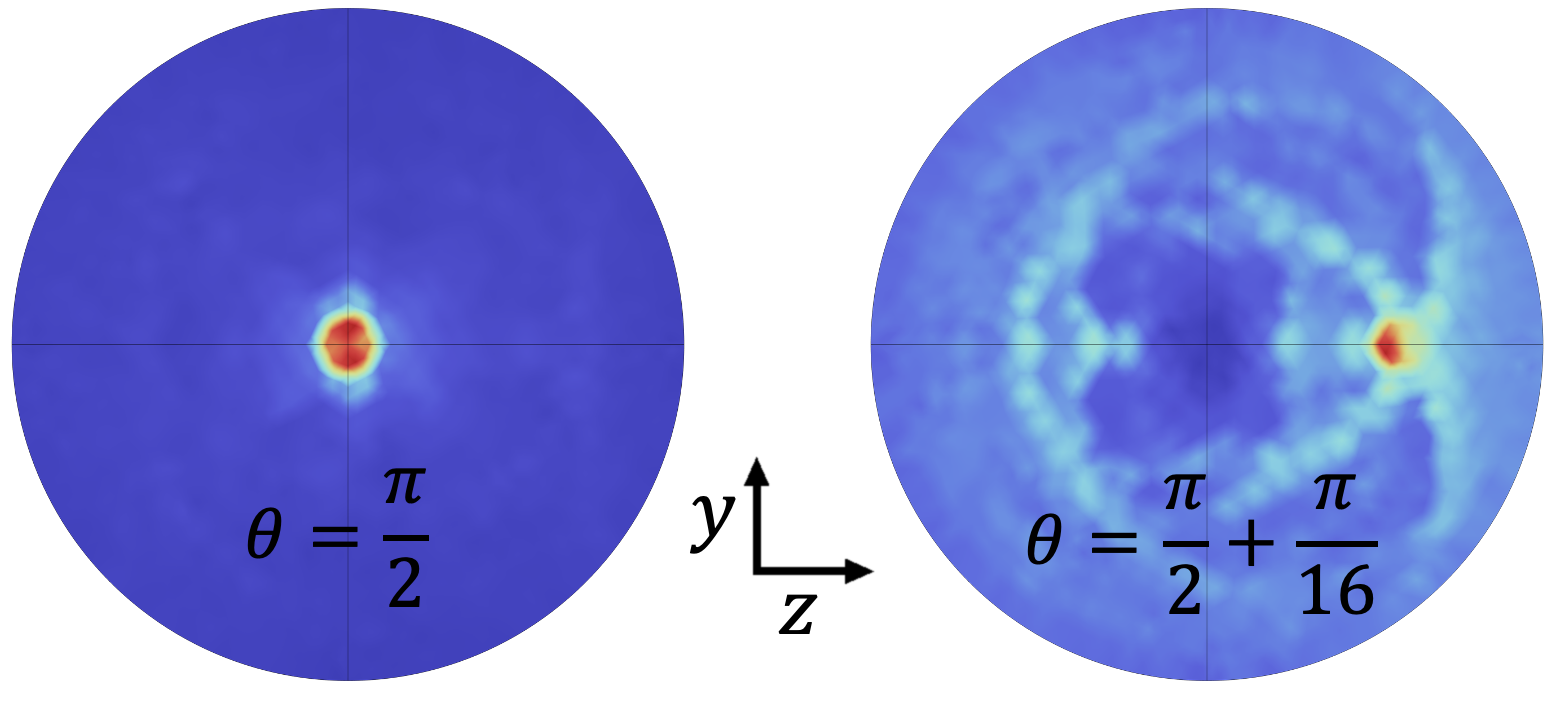}
    \end{subfigure}%
    \vspace{0.1 cm}
    \begin{subfigure}{0.4\textwidth}
        \centering
        \includegraphics[width=\textwidth]{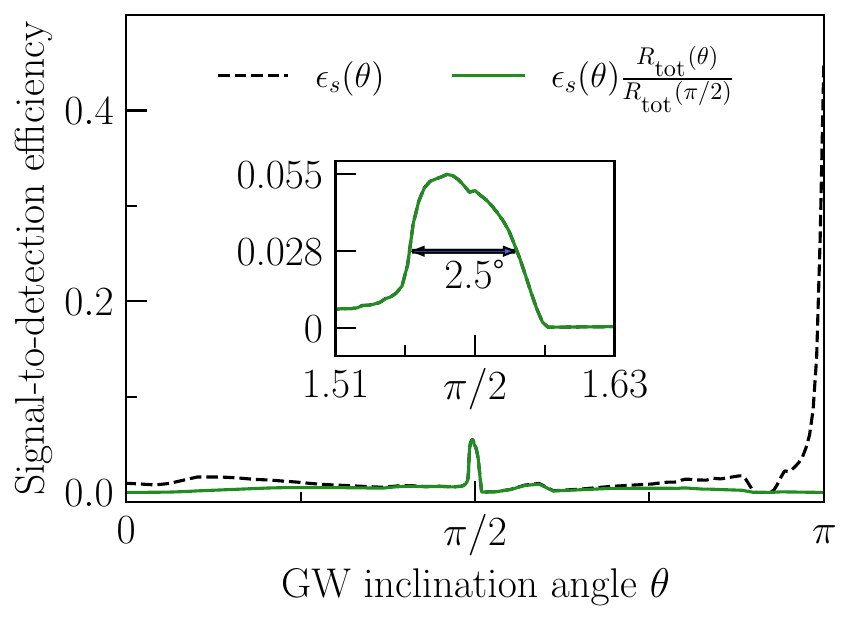}
    \end{subfigure}    
    \caption{\textbf{Top:} Density of incoming rays for two values of $\theta=\text{arctan}(\hat{\bm{k}}_2\cdot\hat{\bm{z}}/\hat{\bm{k}}_2\cdot\hat{\bm{x}})$ at the focal plane of the BREAD cylinder for photons traveling along $\hat{\bm z}$ direction. Red marks a high density and blue a low density. \textbf{Bottom:} The signal-to-detection efficiency as measured in the simulation on its own and weighed with the signal photon rate. The inset shows the focal peak where the bandwidth has been highlighted.}
    \label{fig:focusing}
\end{figure}

The results from our simulation are shown in Fig. \ref{fig:focusing}. The upper panels illustrate the photon heatmap at the top of the cylinder from the outside, where we visualize the focusing of the incoming rays for two benchmark points of $\theta=\frac\pi2$ and $=\frac\pi2+\frac\pi{16}$. At the left, the rays reach BREAD orthogonal to the magnetic field, and the rays get highly focused at the center of the disc, right where the detector is located. To the right, we see how if the HFGW arrives at an angle away from the $\hat{\bm{z}}$ axis, the focal point moves away from the detector. This is nothing but the fact that BREAD is optimized for axion DM searches, for which the signal photons get focused at the detector with high efficiency.

In the bottom panel of Fig.~\ref{fig:focusing} we see a quantitative version of what the top panel illustrates. A sharp peak appears for $\theta\approx\pi/2$ and at $\theta = \pi$. The former is lower because only the fraction of rays which encounter the parabolic mirror can get deflected upwards. At $\theta=\pi$, all rays hit the mirror and can potentially be focused. However, parallel to the static background field, GWs can not convert into photons due to angular momentum conservation and the signal rate $\epsilon(\pi)R_\text{tot}(\pi)=0$. Overall we find $\bar{\epsilon}_s\simeq0.004$. However, every receiver will in reality only be able to read out some fraction ($\sim0.5$ for GigaBREAD \cite{Knirck:2018ojz}) of all rays near the focus. Since we are neglecting these details in our analysis, we use $\bar{\epsilon}_s\simeq10^{-3}$ for our following sensitivity analysis. The bandwidth of the focus around $\pi/2$ is given by $\Delta\theta=0.044$ radians when defined as full width at half maximum.

To put these results into perspective, one of the 8 magnetic coils envisioned for the IAXO experiment \cite{Armengaud:2014gea} has a sensitive angular diameter $\Delta\theta_\text{IAXO}\sim R/L\approx0.04$ radians. Unlike in a coaxial dish antenna, there is no sensitivity to incoming rays along a full $2\pi\Delta\theta\,\text{rad}^2$ solid angle. Therefore, even if all rays within the angular diameter are focused perfectly $\epsilon_\text{IAXO}\approx1$, the sensitive sky fraction is only around $\bar{\epsilon}_\text{IAXO}\sim4\cdot10^{-4}$ \cite{PhysRevLett.132.131402, Aggarwal:2025noe} in our language. Similar angular sensitivity can be found for related axion detectors like CAST or ALPS \cite{PhysRevLett.132.131402}, making the focusing of a BREAD design competitive, despite its non-ideal focusing. Furthermore, the  spherical dish antenna originally proposed in \cite{Horns:2012jf} is also slightly less suited for GW detection than the coaxial BREAD design, as it also only focuses rays from one distinct direction.

\section{Results}
\label{sec:results} 
Combining the results from sections \ref{sec:calculations} and \ref{sec:focusing}, we can estimate the sensitivity that coaxial dish antennas like BREAD can have to gravitational waves. For this purpose, we need to distinguish between different kinds of gravitational wave signals and the type of readout system employed by the dish antenna. In the following we discuss two systems, RF antennas and single photon detectors (SPDs), and we consider the case of a point-like source emitting GW at a particular frequency (or a range of frequencies) and the case of a GW background with a given spectral energy density.

\subsection*{RF Antennas}

A detector for $\sim\text{GHz}$ signals can use a RF antenna to read out the EM fields near the focal point. This system has been realized in the GigaBREAD detector \cite{Knirck:2018ojz}. A benefit, as opposed to using photon detection, is that the readout system receives detailed spectral information on the signal, which can be characterized by the signal power spectral density (PSD)
\begin{equation}\label{eq:signal_PSD}
    S_\text{sig}(\omega_g)=\frac{\omega_g}{\Delta\omega_g}\bar{\epsilon}_sR_\text{tot}^\text{max}\simeq \frac{1}{4}\bar{\epsilon}_s\omega_g^2B^2VL_\text{eff}S_h(\omega_g)\,,
\end{equation}
where $L_\text{eff}$ is the average length the GW traverses within the BREAD detector and $S_h$ is the PSD of the GW signal. Furthermore, $\Delta\omega_g=2\pi \Delta f_g$ is the bandwidth of the GW assuming a frequency independent response of the detector. A benefit of dish antennas is their broadband response where typically $\Delta\omega\sim\mathcal{O}(\omega_g)$. The noise PSD is usually formulated using an effective system temperature $S_\text{noise}(\omega)\simeq T_\text{sys}$. The first GigaBREAD axion search at room temperature was performed with $T_\text{sys}\sim400\,\text{K}-600\,\text{K}$ \cite{GigaBREAD_Axion_search}.

The signal to noise ratio (SNR) in the presence of gaussian thermal fluctuations is given by the Dicke radiometer equation \cite{Dicke} 
\begin{equation}\label{eq:Dicke_Ratiometer_SNR}
    \text{SNR}=\sqrt{t_\text{int}\Delta f_\text{det}}\frac{S_\text{sig}(\omega)}{S_\text{noise}(\omega)}\,, 
\end{equation}
where $t_\text{int}$ is the integration time of the experiment and $\Delta f_\text{det}$ is the measured frequency interval whose upper bound is the frequency range over which $S_\text{sig}(\omega)/S_\text{noise}(\omega)$ is approximately constant.
If the GW signal is persistent and monochromatic, the signal PSD scales as $S_h(\omega_g)\sim h_0^2/\Delta\omega_\text{min}$ where $\Delta\omega_\text{min}$ is the minimal resolvable frequency difference from the detector $\gtrsim 2\pi t_\text{int}^{-1}$, which we assume to be larger than the bandwidth of the GW signal. GigaBREAD's axion search chose $\omega_\text{min}/2\pi=7.8\,\text{kHz}$ as this approximately matches the axion bandwidth \cite{GigaBREAD_Axion_search}. Using this, the minimal detectable GW strain in the presence of thermal noise becomes for $\Delta f_\text{det}=\Delta f_\text{min}$
\begin{align}\label{eq:h_min_th}
    h_0^\text{th}\,& = 7\cdot10^{-18}\left(\frac{\Delta f_\text{min}}{7.8\,\text{kHz}}\frac{30\,\text{days}}{t_\text{int}}\right)^{\frac{1}{4}} \nonumber\\ &\left(\frac{\text{SNR}}{5}\frac{10^{-4}}{\bar{\epsilon}_s}\frac{T_\text{sys}}{600\,\text{K}}\frac{0.017\,\text{m}^4}{L_\text{eff}V}\right)^{\frac{1}{2}}\frac{10\,\text{GHz}}{f_g}\frac{3.9\,\text{T}}{B}\,,
\end{align}
where GigaBREAD parameters have been inserted for reference. A discussion on the choice of $\bar{\epsilon}_s$ can be found in appendix \ref{sec:low_f_focusing}. Note that while the integration time is assumed to be 30 days, a localized monochromatic source will only remain focused for the fraction $\sim2.5^\circ/180^\circ$ of a day, assuming the focal width shown in Fig. \ref{fig:focusing}. However, this reduction is already accounted for by $\bar{\epsilon}_s$.

\begin{figure}[t]
    \centering
    \includegraphics[width=0.48\textwidth]{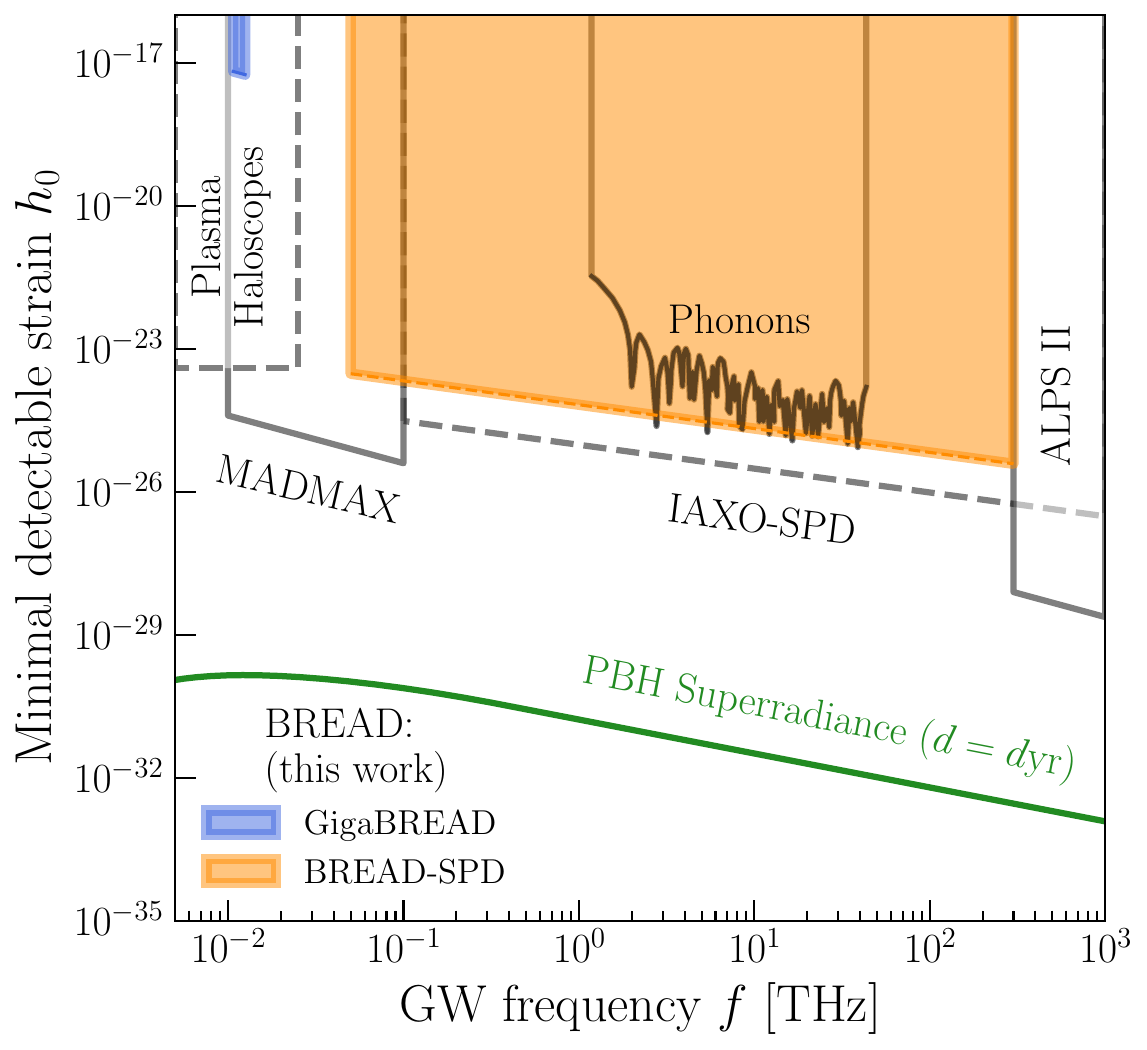}
    \caption{Sensitivity of coaxial dish antennas to monochromatic gravitational waves compared with other operating and proposed experiments in a similar frequency range. Dashed lines are used when more than one detector setup is needed to cover the entire frequency range. For reference, the signal of PBH superradiance is shown. Details on the assumptions can be found in section \ref{sec:results}.}
    \label{fig:hminbound}
\end{figure}

The GigaBREAD sensitivity of point-like GW sources is shown in blue at the top-left corner of Fig.~\ref{fig:hminbound}. We can see how this setup is not particularly well suited for GW searches. For comparisons the figure also shows the reach of Plasma Haloscopes like ALPHA~\cite{Capdevilla:2024cby} and Dielectric Haloscopes like MADMAX~\cite{Domcke:2024eti} which can operate in a frequency range that overlaps with GigaBREAD. A potential source of GWs at these frequencies includes PBH superradiance \cite{Arvanitaki:2009fg, Arvanitaki:2010sy, Yoshino:2013ofa}, shown as a red line. The strain corresponds to a source located at a distance $d_\text{yr}$, defined as the distance where one event per year is expected, assuming that PBHs make up nearly all DM in the Universe.


\begin{figure}[t]
    \centering
    \includegraphics[width=0.48\textwidth]{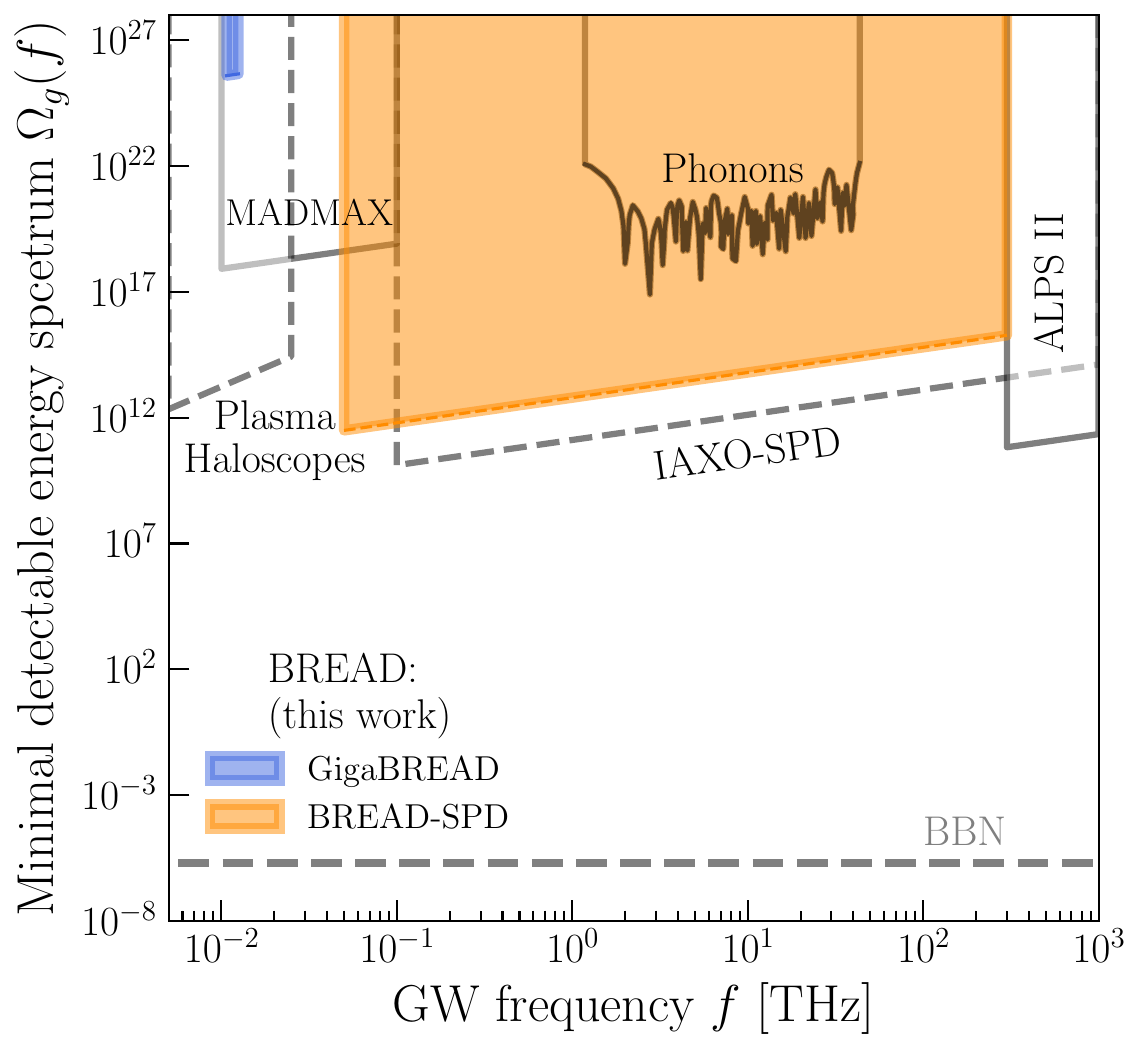}
    \caption{Sensitivity of coaxial dish antennas to stochastic gravitational wave backgrounds compared with other operating and proposed experiments in a similar frequency range. Dashed lines are used when more than one detector setup is needed to cover the entire frequency range. The experimental upper bound from BBN is shown for reference. Details on the assumptions can be found in section \ref{sec:results}.}
    \label{fig:Omegaminbound}
\end{figure}

In Fig~\ref{fig:Omegaminbound}, we evaluate the sensitivity to stochastic gravitational wave backgrounds, which typically have a broad frequency range characterized by the spectral energy density $\Omega_g(f)=4\pi/(3H_0^2)f^3S_h(f)$, where $H_0$ is the Hubble constant \cite{Aggarwal:2025noe}. Using Eqs.~\eqref{eq:signal_PSD} and \eqref{eq:Dicke_Ratiometer_SNR} we can infer the minimal detectable energy spectrum
\begin{align}\label{eq:Omega_min_th}
    \Omega_g^\text{th}(f)&=3.6\cdot10^{25}\left(\frac{2\,\text{GHz}}{\Delta f_\text{det}}\frac{30\,\text{days}}{t_\text{int}}\right)^{\frac{1}{2}}\frac{\text{SNR}}{5}\\ &\frac{10^{-4}}{\bar{\epsilon}_s}\frac{T_\text{sys}}{600\,\text{K}}\frac{0.017\,\text{m}^4}{L_\text{eff}V}\frac{f_g}{10\,\text{GHz}}\left(\frac{3.9\,\text{T}}{B}\right)^2\,,
\end{align}
where the sensitivity now increases by using the largest possible bandwidth $\Delta f_\text{det}<\Delta f_g$. The GigaBREAD sensitivity on GW backgrounds is shown as a blue band at the top-left corner of Fig.~\ref{fig:Omegaminbound}. As can be clearly seen, the $\Omega_g$ sensitivity of this setup is still far from being competitive with other proposed experiments like  ALPHA and MADMAX.

\subsection*{Single Photon Detectors}


A significant improvement in sensitivity can be reached if a BREAD-type detector is built in a frequency range where single photon detectors (SPDs) can be used for the signal readout~\cite{Verma:2020gso}. State-of-the art SPDs achieve dark count rates as low as $R_D\gtrsim10^{-4}$ Hz \cite{Wollman:17} at ultraviolet frequencies and can thus operate well below the standard quantum limit applying to linear quadrature measurements $R_D^\text{SQL}/\Delta f\geq1$ \cite{PhysRevD.88.035020}. However, much progress is also being made on achieving comparable dark counts at frequencies as low as $\sim10\,\text{GHz}$ \cite{Braggio:2024xed}. Another possible detector technology in the THz range are bolometers \cite{BREAD:2021tpx}. Since their noise performance is expected to be slightly worse than future SPDs, we are not considering their sensitivity explicitly here. 

The SNR for single photon detection is given by \cite{BREAD:2021tpx}
\begin{equation}
    \text{SNR}=\sqrt{t_\text{int}}\frac{R_\text{sig}}{\sqrt{R_D}}\,.
\end{equation}
The resulting minimal detectable monochromatic strain then becomes
\begin{align}\label{eq:h_min_SPD}
    h_0^\text{SPD} =\,& 6.7\cdot10^{-25}\left(\frac{R_D}{10^{-4}\,\text{Hz}}\frac{1\,\text{yr}}{t_\text{int}}\right)^{\frac{1}{4}}\\ &\left(\frac{\text{SNR}}{5}\frac{10^{-3}}{\bar{\epsilon}_s}\frac{1\,\text{m}^4}{L_\text{eff}V}\frac{\text{THz}}{f_g}\right)^{\frac{1}{2}}\frac{10\,\text{T}}{B}\,,
\end{align}
where there is no dependence on $\Delta f_\text{min}$ anymore, since the frequency range of photons measured by the photon counter is implicit in $R_D$ and can not be adjusted as for the spectral readout considered before. Furthermore, the minimal detectable spectral fraction of GW energy in the universe is given by
\begin{align}\label{eq:Omega_min_SPD}
    \Omega_g^\text{SPD}(f)&=6.2\cdot10^{12}\left(\frac{R_D}{10^{-4}\,\text{Hz}}\frac{1\,\text{yr}}{t_\text{int}}\right)^{\frac{1}{2}}\frac{\text{SNR}}{5}\\ &\frac{10^{-3}}{\bar{\epsilon}_s}\frac{1\,\text{m}^4}{L_\text{eff}V}\frac{f_g}{\text{THz}}\frac{f_g}{\Delta f}\left(\frac{10\,\text{T}}{B}\right)^2\,.
\end{align}
The exact bandwidth of a single experiment depends on the specific SPD used but is expected to be $\Delta f\sim f_g$. Therefore, it is also possible that more than one experimental setup would be needed to explore the whole frequency range shown in Figs. \ref{fig:hminbound} and \ref{fig:Omegaminbound}. In both cases, a slightly improved signal rate based on the assumptions in Ref.~\cite{BREAD:2021tpx} along with the low noise background, has led to a significant improvement in sensitivity. Figs.~\ref{fig:hminbound} and \ref{fig:Omegaminbound} show the sensitivity of the upgraded SPD detector, using the benchmark parameters from Eqs. \eqref{eq:h_min_SPD} and \eqref{eq:Omega_min_SPD}.
A future BREAD-type detector can set highly competitive GW limits in comparison to proposed and operating experiments in this frequency range. This includes low frequency IAXO \cite{Armengaud:2014gea, Ringwald:2020ist}, ALPS II \cite{Bahre:2013ywa}, single-phonon excitations \cite{Kahn:2023mrj}, dielectric haloscopes like MADMAX \cite{MADMAX:2019pub,Domcke:2024eti} and plasma haloscopes \cite{Capdevilla:2024cby}.\\

\section{Conclusions}
\label{sec:conclusion}

We have estimated the sensitivity of the Broadband Reflector Experiment for Axion Detection (BREAD) to high-frequency gravitational waves (HFGWs). This experiment is designed to detect photons that are produced when ambient axion dark matter scatters off the $B$ field in the system. BREAD's focusing system is optimized for DM searches, where the signal photons are produced perpendicular to the walls of the cylindrical cavity. At the desired high frequencies, GWs convert following the direction of arrival, which in principle suggests a disadvantage for BREAD regarding GW detection. However, while axion-to-photon conversion occurs at the boundary, the GW-to-photon conversion takes place throughout the entire volume, leading to a sizable effect. In essence, DM searches at BREAD benefit from the focusing system while GWs searches benefit from the large volume, which makes BREAD a competitive setup for both searches.

Our results show that GW signals can generate a measurable photon conversion rate inside the BREAD, though practical detection depends on the specific characteristics and sensitivities of the photosensor used. By evaluating signal-to-noise ratio (SNR) capabilities for two distinct detectors suitable for $\sim$GHz and $\sim$THz respectively, we determine the minimal detectable GW signals in Figs.~\ref{fig:hminbound} and \ref{fig:Omegaminbound}. Single photon detectors in BREAD (which we call BREAD-SPD) will be able to probe the high-frequency range from 0.05 to 200 THz whereas RF antennas can be used to detect photons with $\sim 10$GHz.

At radio frequencies, GigaBREAD provides weaker bounds \( h_0 \sim 10^{-17} \) far away from the other proposed experiments.
Yet, BREAD-SPD offers enhanced sensitivity in the photon energy band reaching a minimum detectable amplitude of \( h_0 \sim 10^{-25} \). The drastic sensitivity difference originates mainly from the much lower noise of SPD quantum sensors.
This supports the prospective improvement in stochastic GW detection. These findings indicate that the BREAD is capable of exploring uncharted territory in the GW parameter space with similar sensitivities to that of other experiments that probe $\sim$ THz frequencies even with an apparatus relatively \emph{small} compared to competitors which operate at similar frequencies.

Future work may build upon the methodologies presented in this paper, probing optimized sensor configurations and possible SNR enhancement strategies to further advance HFGW detection. Another promising avenue for further studies is the potential detection of the Cosmic Gravitational-wave Background (CGB) at 0.1~THz, an interesting target for HFGW searches.

\section*{Acknowledgments}

We thank Asher Berlin, Tanner Trickle, Wolfram Ratzinger, Pedro Schwaller, Sebastian Schenk, Alex Millar, Andrew Sonnenschein, Scott Watson, Sebastian Ellis, and Joachim Kopp for wonderful discussions.
This manuscript has been authored by Fermi Forward Discovery Group, LLC under Contract No. 89243024CSC000002 with the U.S. Department of Energy, Office of Science, Office of High Energy Physics. 
It is partially supported by this office's the Quantum Information Science Enabled Discovery (QuantISED) program.
RH is also supported by the U.S. Department of Energy, Office of Science, National Quantum Information Science Research Centers, Superconducting Quantum Materials and Systems (SQMS) Center under the contract No. DE-AC02-07CH11359.
T. Krokotsch acknowledges support by the Deutsche Forschungsgemeinschaft (DFG, German Research Foundation) under Germany's Excellence Strategy - EXC 2121 `Quantum Universe' - 390833306 and by the European Union?s Horizon Europe Marie Sklodowska-Curie Staff Exchanges programme under grant agreement no. 101086276.

\appendix
\section{Focusing at radio frequencies}
\label{sec:low_f_focusing}

For GW frequencies $\sim\text{GHz}$ the gravitational wavelength is comparable to the size of the detector and ray optics are not a valid approximation. This is because interference and diffraction effects become relevant and resonances can be excited. Instead, we need to solve Maxwell's equations in the volume of the detector, find the E and B fields, and from them compute the GW power converted into electromagnetic power. Varying the action, Eq.~\ref{eq:action}, with respect to the field $A^\mu$ one can obtain the inhomogeneous Maxwell's equations
\begin{equation}
\partial_\mu F^{\mu\nu} = -j^\nu_{\rm eff},
\label{eq:Maxwell_covariant}
\end{equation}
where we have defined a gravitationally-induced effective current
\begin{equation}\label{eq:jeff}
    j^\mu_\text{eff}\equiv \partial_\nu\left( \frac12 h\,F^{\mu\nu}+h^\nu_{~~\alpha}F^{\alpha\mu} -h^\mu_{~~\alpha}F^{\alpha\nu} \right)
\end{equation}
This effective current acts as an additional source term for the E and B fields inside the detector volume
\begin{align}
    \nabla\cdot\textbf{E}=\rho + \rho_\text{eff}\\
    \nabla\times\textbf{B}-\partial_t\textbf{E}=\bm{j}+\bm{j}_\text{eff}
\end{align}

In order to investigate whether the BREAD geometry can still focus electromagnetic waves from radio frequency GWs, we use COMSOL to solve these equations, assuming a monochromatic GW traveling orthogonal to the magnetic field. In the $\sim\text{GHz}$ frequency range, we need to use the complete resumed expression of $\bm{J}_\text{eff}$ derived in \cite{Berlin:2021txa} for which the remaining experiment is approximately static with respect to the laboratory frame. Furthermore, the experiment is not axially symmetric as for an axion current and a full 3D simulation needs to be performed. The top of the BREAD cylinder is taken to perfectly absorb RF power, while the walls are assumed to be perfect conductors. The cylinder was assumed to have radius $R_\text{top}=0.2\,\text{m}$, and height $L=0.57\,\text{m}$ and the flat part of the inner reflector a radius $R_\text{flat}=15\,\text{mm}$. 

\begin{figure}[t]
    \centering
    \includegraphics[width=0.48\textwidth]{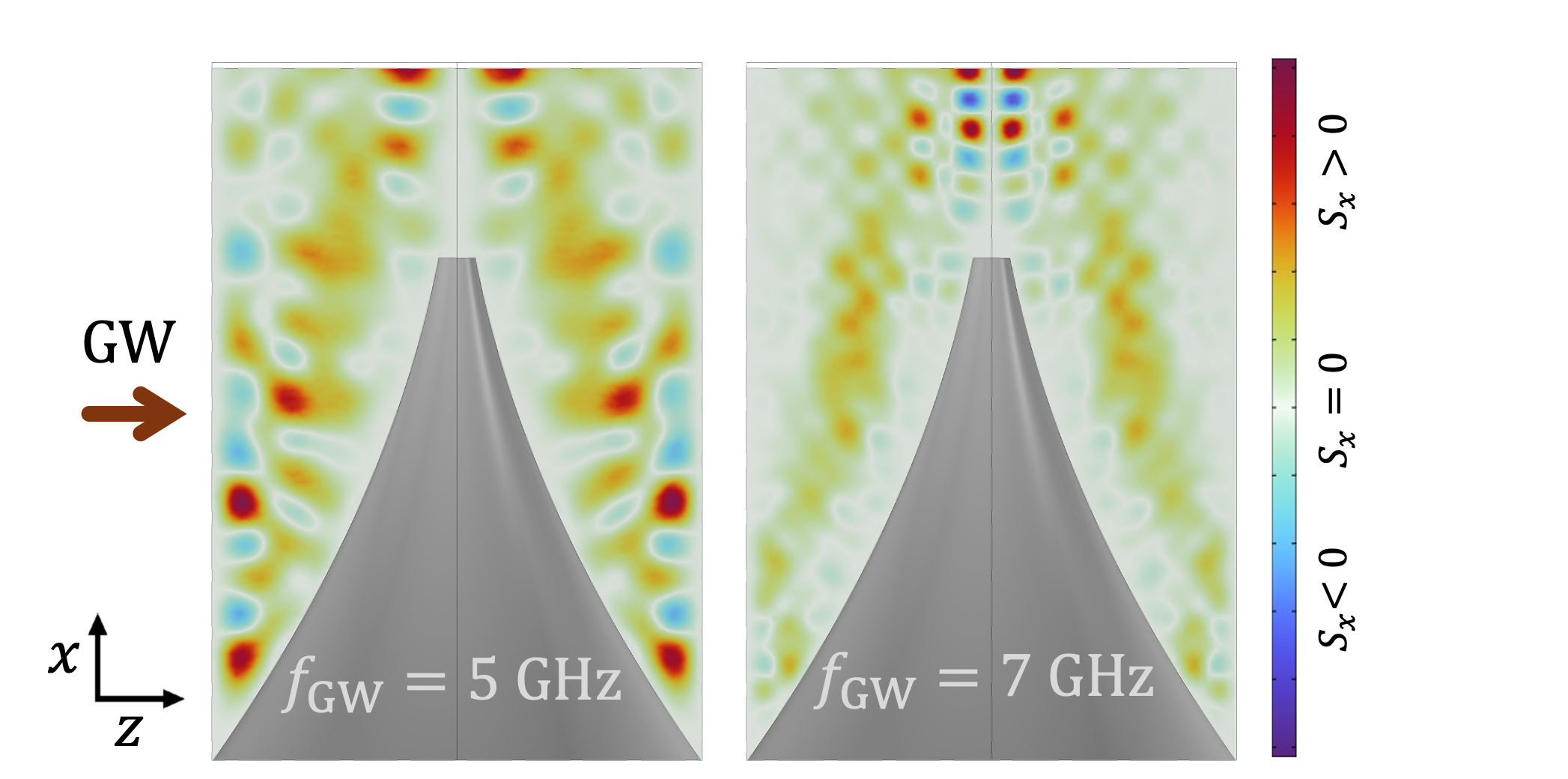}
    \caption{Full field simulation of the Poynting flux in $x$ direction $S_x$ as sourced by the effective current density from a monochromatic gravitational wave in $z$ direction in the proper detector frame.}
    \label{fig:low_f_focus}
\end{figure}
Therefore, the GigaBREAD sensitivity in Figs. \ref{fig:hminbound} and \ref{fig:Omegaminbound} were drawn assuming conservatively $\bar{\epsilon}_s=10^{-4}$ and should not be taken as safe exclusion limits. An experimental estimate of the sensitivity could be obtained by measuring the distribution of the reflected electric field injected by the RF receiver and calculating its overlap with $j_\text{eff}^i$ as described in \cite{Egge_2023, Egge_2024}.

Fig. \ref{fig:low_f_focus} illustrates the model and shows the resulting Poynting flow $S_x$ along the symmetry axis of the cylinder. Red color marks an upward power flow, possibly towards a detector and blue marks power flowing in the opposite direction. While for most simulated GW frequencies like $f_\text{GW}=7\,\text{GHz}$, focal points can be found above the parabolic mirror, exceptions like at  $f_\text{GW}=7\,\text{GHz}$ occur as well. Most likely, this is due to unwanted standing waves forming in the cylinder which do not support large fields at the desired focal point. Consequently, the focusing of GW signals for $f_\text{GW}\lesssim\,\text{GHz}$ depends strongly on the frequency and details on the experimental setup. Therefore, we conclude that BREAD antennas are not well suited for GW detection in this frequency range but can in principle still pick up signals. The first axion search with the GigaBREAD detector \cite{GigaBREAD_Axion_search} was performed around $10\,\text{GHz}$ and thus not in a range where we can safely assume ray optics.

\bibliography{Ref}

\begin{thebibliography}{72}%
\makeatletter
\providecommand \@ifxundefined [1]{%
 \@ifx{#1\undefined}
}%
\providecommand \@ifnum [1]{%
 \ifnum #1\expandafter \@firstoftwo
 \else \expandafter \@secondoftwo
 \fi
}%
\providecommand \@ifx [1]{%
 \ifx #1\expandafter \@firstoftwo
 \else \expandafter \@secondoftwo
 \fi
}%
\providecommand \natexlab [1]{#1}%
\providecommand \enquote  [1]{``#1''}%
\providecommand \bibnamefont  [1]{#1}%
\providecommand \bibfnamefont [1]{#1}%
\providecommand \citenamefont [1]{#1}%
\providecommand \href@noop [0]{\@secondoftwo}%
\providecommand \href [0]{\begingroup \@sanitize@url \@href}%
\providecommand \@href[1]{\@@startlink{#1}\@@href}%
\providecommand \@@href[1]{\endgroup#1\@@endlink}%
\providecommand \@sanitize@url [0]{\catcode `\\12\catcode `\$12\catcode
  `\&12\catcode `\#12\catcode `\^12\catcode `\_12\catcode `\%12\relax}%
\providecommand \@@startlink[1]{}%
\providecommand \@@endlink[0]{}%
\providecommand \url  [0]{\begingroup\@sanitize@url \@url }%
\providecommand \@url [1]{\endgroup\@href {#1}{\urlprefix }}%
\providecommand \urlprefix  [0]{URL }%
\providecommand \Eprint [0]{\href }%
\providecommand \doibase [0]{https://doi.org/}%
\providecommand \selectlanguage [0]{\@gobble}%
\providecommand \bibinfo  [0]{\@secondoftwo}%
\providecommand \bibfield  [0]{\@secondoftwo}%
\providecommand \translation [1]{[#1]}%
\providecommand \BibitemOpen [0]{}%
\providecommand \bibitemStop [0]{}%
\providecommand \bibitemNoStop [0]{.\EOS\space}%
\providecommand \EOS [0]{\spacefactor3000\relax}%
\providecommand \BibitemShut  [1]{\csname bibitem#1\endcsname}%
\let\auto@bib@innerbib\@empty
\bibitem [{\citenamefont {Abbott}\ \emph {et~al.}(2016)\citenamefont {Abbott}
  \emph {et~al.}}]{LIGOScientific:2016aoc}%
  \BibitemOpen
  \bibfield  {author} {\bibinfo {author} {\bibfnamefont {B.~P.}\ \bibnamefont
  {Abbott}} \emph {et~al.} (\bibinfo {collaboration} {LIGO Scientific,
  Virgo}),\ }\bibfield  {title} {\bibinfo {title} {{Observation of
  Gravitational Waves from a Binary Black Hole Merger}},\ }\href
  {https://doi.org/10.1103/PhysRevLett.116.061102} {\bibfield  {journal}
  {\bibinfo  {journal} {Phys. Rev. Lett.}\ }\textbf {\bibinfo {volume} {116}},\
  \bibinfo {pages} {061102} (\bibinfo {year} {2016})},\ \Eprint
  {https://arxiv.org/abs/1602.03837} {arXiv:1602.03837 [gr-qc]} \BibitemShut
  {NoStop}%
\bibitem [{\citenamefont {Aggarwal}\ \emph {et~al.}(2021)\citenamefont
  {Aggarwal} \emph {et~al.}}]{Aggarwal:2020olq}%
  \BibitemOpen
  \bibfield  {author} {\bibinfo {author} {\bibfnamefont {N.}~\bibnamefont
  {Aggarwal}} \emph {et~al.},\ }\bibfield  {title} {\bibinfo {title}
  {{Challenges and opportunities of gravitational-wave searches at MHz to GHz
  frequencies}},\ }\href {https://doi.org/10.1007/s41114-021-00032-5}
  {\bibfield  {journal} {\bibinfo  {journal} {Living Rev. Rel.}\ }\textbf
  {\bibinfo {volume} {24}},\ \bibinfo {pages} {4} (\bibinfo {year} {2021})},\
  \Eprint {https://arxiv.org/abs/2011.12414} {arXiv:2011.12414 [gr-qc]}
  \BibitemShut {NoStop}%
\bibitem [{\citenamefont {Aggarwal}\ \emph {et~al.}(2025)\citenamefont
  {Aggarwal} \emph {et~al.}}]{Aggarwal:2025noe}%
  \BibitemOpen
  \bibfield  {author} {\bibinfo {author} {\bibfnamefont {N.}~\bibnamefont
  {Aggarwal}} \emph {et~al.},\ }\bibfield  {title} {\bibinfo {title}
  {{Challenges and Opportunities of Gravitational Wave Searches above 10
  kHz}},\ }\href@noop {} {\  (\bibinfo {year} {2025})},\ \Eprint
  {https://arxiv.org/abs/2501.11723} {arXiv:2501.11723 [gr-qc]} \BibitemShut
  {NoStop}%
\bibitem [{\citenamefont {McDonald}\ and\ \citenamefont
  {Ellis}(2024)}]{McDonald:2024nxj}%
  \BibitemOpen
  \bibfield  {author} {\bibinfo {author} {\bibfnamefont {J.~I.}\ \bibnamefont
  {McDonald}}\ and\ \bibinfo {author} {\bibfnamefont {S.~A.~R.}\ \bibnamefont
  {Ellis}},\ }\bibfield  {title} {\bibinfo {title} {{Resonant conversion of
  gravitational waves in neutron star magnetospheres}},\ }\href
  {https://doi.org/10.1103/PhysRevD.110.103003} {\bibfield  {journal} {\bibinfo
   {journal} {Phys. Rev. D}\ }\textbf {\bibinfo {volume} {110}},\ \bibinfo
  {pages} {103003} (\bibinfo {year} {2024})},\ \Eprint
  {https://arxiv.org/abs/2406.18634} {arXiv:2406.18634 [hep-ph]} \BibitemShut
  {NoStop}%
\bibitem [{\citenamefont {Garcia-Cely}\ and\ \citenamefont
  {Ringwald}(2024)}]{Garcia-Cely:2024ujr}%
  \BibitemOpen
  \bibfield  {author} {\bibinfo {author} {\bibfnamefont {C.}~\bibnamefont
  {Garcia-Cely}}\ and\ \bibinfo {author} {\bibfnamefont {A.}~\bibnamefont
  {Ringwald}},\ }\bibfield  {title} {\bibinfo {title} {{Complete
  Gravitational-Wave Spectrum of the Sun}},\ }\href@noop {} {\  (\bibinfo
  {year} {2024})},\ \Eprint {https://arxiv.org/abs/2407.18297}
  {arXiv:2407.18297 [hep-ph]} \BibitemShut {NoStop}%
\bibitem [{\citenamefont {Ghiglieri}\ and\ \citenamefont
  {Laine}(2015)}]{Ghiglieri:2015nfa}%
  \BibitemOpen
  \bibfield  {author} {\bibinfo {author} {\bibfnamefont {J.}~\bibnamefont
  {Ghiglieri}}\ and\ \bibinfo {author} {\bibfnamefont {M.}~\bibnamefont
  {Laine}},\ }\bibfield  {title} {\bibinfo {title} {{Gravitational wave
  background from Standard Model physics: Qualitative features}},\ }\href
  {https://doi.org/10.1088/1475-7516/2015/07/022} {\bibfield  {journal}
  {\bibinfo  {journal} {JCAP}\ }\textbf {\bibinfo {volume} {07}},\ \bibinfo
  {pages} {022}},\ \Eprint {https://arxiv.org/abs/1504.02569} {arXiv:1504.02569
  [hep-ph]} \BibitemShut {NoStop}%
\bibitem [{\citenamefont {Giovannini}(2020)}]{Giovannini:2019oii}%
  \BibitemOpen
  \bibfield  {author} {\bibinfo {author} {\bibfnamefont {M.}~\bibnamefont
  {Giovannini}},\ }\bibfield  {title} {\bibinfo {title} {{Primordial
  backgrounds of relic gravitons}},\ }\href
  {https://doi.org/10.1016/j.ppnp.2020.103774} {\bibfield  {journal} {\bibinfo
  {journal} {Prog. Part. Nucl. Phys.}\ }\textbf {\bibinfo {volume} {112}},\
  \bibinfo {pages} {103774} (\bibinfo {year} {2020})},\ \Eprint
  {https://arxiv.org/abs/1912.07065} {arXiv:1912.07065 [astro-ph.CO]}
  \BibitemShut {NoStop}%
\bibitem [{\citenamefont {Ghiglieri}\ \emph {et~al.}(2020)\citenamefont
  {Ghiglieri}, \citenamefont {Jackson}, \citenamefont {Laine},\ and\
  \citenamefont {Zhu}}]{Ghiglieri:2020mhm}%
  \BibitemOpen
  \bibfield  {author} {\bibinfo {author} {\bibfnamefont {J.}~\bibnamefont
  {Ghiglieri}}, \bibinfo {author} {\bibfnamefont {G.}~\bibnamefont {Jackson}},
  \bibinfo {author} {\bibfnamefont {M.}~\bibnamefont {Laine}},\ and\ \bibinfo
  {author} {\bibfnamefont {Y.}~\bibnamefont {Zhu}},\ }\bibfield  {title}
  {\bibinfo {title} {{Gravitational wave background from Standard Model
  physics: Complete leading order}},\ }\href
  {https://doi.org/10.1007/JHEP07(2020)092} {\bibfield  {journal} {\bibinfo
  {journal} {JHEP}\ }\textbf {\bibinfo {volume} {07}},\ \bibinfo {pages}
  {092}},\ \Eprint {https://arxiv.org/abs/2004.11392} {arXiv:2004.11392
  [hep-ph]} \BibitemShut {NoStop}%
\bibitem [{\citenamefont {Ringwald}\ \emph {et~al.}(2021)\citenamefont
  {Ringwald}, \citenamefont {Sch\"utte-Engel},\ and\ \citenamefont
  {Tamarit}}]{Ringwald:2020ist}%
  \BibitemOpen
  \bibfield  {author} {\bibinfo {author} {\bibfnamefont {A.}~\bibnamefont
  {Ringwald}}, \bibinfo {author} {\bibfnamefont {J.}~\bibnamefont
  {Sch\"utte-Engel}},\ and\ \bibinfo {author} {\bibfnamefont {C.}~\bibnamefont
  {Tamarit}},\ }\bibfield  {title} {\bibinfo {title} {{Gravitational Waves as a
  Big Bang Thermometer}},\ }\href
  {https://doi.org/10.1088/1475-7516/2021/03/054} {\bibfield  {journal}
  {\bibinfo  {journal} {JCAP}\ }\textbf {\bibinfo {volume} {03}},\ \bibinfo
  {pages} {054}},\ \Eprint {https://arxiv.org/abs/2011.04731} {arXiv:2011.04731
  [hep-ph]} \BibitemShut {NoStop}%
\bibitem [{\citenamefont {Giovannini}(2023)}]{Giovannini:2023itq}%
  \BibitemOpen
  \bibfield  {author} {\bibinfo {author} {\bibfnamefont {M.}~\bibnamefont
  {Giovannini}},\ }\bibfield  {title} {\bibinfo {title} {{Relic gravitons and
  high-frequency detectors}},\ }\href
  {https://doi.org/10.1088/1475-7516/2023/05/056} {\bibfield  {journal}
  {\bibinfo  {journal} {JCAP}\ }\textbf {\bibinfo {volume} {05}},\ \bibinfo
  {pages} {056}},\ \Eprint {https://arxiv.org/abs/2303.11928} {arXiv:2303.11928
  [gr-qc]} \BibitemShut {NoStop}%
\bibitem [{\citenamefont {Cabral}\ and\ \citenamefont
  {Lobo}(2017)}]{Cabral:2016klm}%
  \BibitemOpen
  \bibfield  {author} {\bibinfo {author} {\bibfnamefont {F.}~\bibnamefont
  {Cabral}}\ and\ \bibinfo {author} {\bibfnamefont {F.~S.~N.}\ \bibnamefont
  {Lobo}},\ }\bibfield  {title} {\bibinfo {title} {{Gravitational waves and
  electrodynamics: New perspectives}},\ }\href
  {https://doi.org/10.1140/epjc/s10052-017-4791-z} {\bibfield  {journal}
  {\bibinfo  {journal} {Eur. Phys. J. C}\ }\textbf {\bibinfo {volume} {77}},\
  \bibinfo {pages} {237} (\bibinfo {year} {2017})},\ \Eprint
  {https://arxiv.org/abs/1603.08157} {arXiv:1603.08157 [gr-qc]} \BibitemShut
  {NoStop}%
\bibitem [{\citenamefont {Garg}\ and\ \citenamefont
  {Dodin}(2022)}]{Garg:2022wdm}%
  \BibitemOpen
  \bibfield  {author} {\bibinfo {author} {\bibfnamefont {D.}~\bibnamefont
  {Garg}}\ and\ \bibinfo {author} {\bibfnamefont {I.~Y.}\ \bibnamefont
  {Dodin}},\ }\bibfield  {title} {\bibinfo {title} {{Gravitational wave modes
  in matter}},\ }\href {https://doi.org/10.1088/1475-7516/2022/08/017}
  {\bibfield  {journal} {\bibinfo  {journal} {JCAP}\ }\textbf {\bibinfo
  {volume} {08}}\bibfield  {number} {\bibinfo  {number} { (08)},\ \bibinfo
  {pages} {017}},\ }\Eprint {https://arxiv.org/abs/2204.09095}
  {arXiv:2204.09095 [gr-qc]} \BibitemShut {NoStop}%
\bibitem [{\citenamefont {Garg}\ and\ \citenamefont
  {Dodin}(2024)}]{Garg:2023yaw}%
  \BibitemOpen
  \bibfield  {author} {\bibinfo {author} {\bibfnamefont {D.}~\bibnamefont
  {Garg}}\ and\ \bibinfo {author} {\bibfnamefont {I.~Y.}\ \bibnamefont
  {Dodin}},\ }\bibfield  {title} {\bibinfo {title} {{Self-consistent
  interaction of linear gravitational and electromagnetic waves in
  non-magnetized plasma}},\ }\href
  {https://doi.org/10.1088/1475-7516/2024/02/045} {\bibfield  {journal}
  {\bibinfo  {journal} {JCAP}\ }\textbf {\bibinfo {volume} {02}},\ \bibinfo
  {pages} {045}},\ \Eprint {https://arxiv.org/abs/2307.05844} {arXiv:2307.05844
  [gr-qc]} \BibitemShut {NoStop}%
\bibitem [{\citenamefont {Dolgov}\ and\ \citenamefont
  {Ejlli}(2011)}]{Dolgov:2011cq}%
  \BibitemOpen
  \bibfield  {author} {\bibinfo {author} {\bibfnamefont {A.~D.}\ \bibnamefont
  {Dolgov}}\ and\ \bibinfo {author} {\bibfnamefont {D.}~\bibnamefont {Ejlli}},\
  }\bibfield  {title} {\bibinfo {title} {{Relic gravitational waves from light
  primordial black holes}},\ }\href
  {https://doi.org/10.1103/PhysRevD.84.024028} {\bibfield  {journal} {\bibinfo
  {journal} {Phys. Rev. D}\ }\textbf {\bibinfo {volume} {84}},\ \bibinfo
  {pages} {024028} (\bibinfo {year} {2011})},\ \Eprint
  {https://arxiv.org/abs/1105.2303} {arXiv:1105.2303 [astro-ph.CO]}
  \BibitemShut {NoStop}%
\bibitem [{\citenamefont {Franciolini}\ \emph {et~al.}(2022)\citenamefont
  {Franciolini}, \citenamefont {Maharana},\ and\ \citenamefont
  {Muia}}]{Franciolini:2022htd}%
  \BibitemOpen
  \bibfield  {author} {\bibinfo {author} {\bibfnamefont {G.}~\bibnamefont
  {Franciolini}}, \bibinfo {author} {\bibfnamefont {A.}~\bibnamefont
  {Maharana}},\ and\ \bibinfo {author} {\bibfnamefont {F.}~\bibnamefont
  {Muia}},\ }\bibfield  {title} {\bibinfo {title} {{Hunt for light primordial
  black hole dark matter with ultrahigh-frequency gravitational waves}},\
  }\href {https://doi.org/10.1103/PhysRevD.106.103520} {\bibfield  {journal}
  {\bibinfo  {journal} {Phys. Rev. D}\ }\textbf {\bibinfo {volume} {106}},\
  \bibinfo {pages} {103520} (\bibinfo {year} {2022})},\ \Eprint
  {https://arxiv.org/abs/2205.02153} {arXiv:2205.02153 [astro-ph.CO]}
  \BibitemShut {NoStop}%
\bibitem [{\citenamefont {Giudice}\ \emph {et~al.}(2016)\citenamefont
  {Giudice}, \citenamefont {McCullough},\ and\ \citenamefont
  {Urbano}}]{Giudice:2016zpa}%
  \BibitemOpen
  \bibfield  {author} {\bibinfo {author} {\bibfnamefont {G.~F.}\ \bibnamefont
  {Giudice}}, \bibinfo {author} {\bibfnamefont {M.}~\bibnamefont
  {McCullough}},\ and\ \bibinfo {author} {\bibfnamefont {A.}~\bibnamefont
  {Urbano}},\ }\bibfield  {title} {\bibinfo {title} {{Hunting for Dark
  Particles with Gravitational Waves}},\ }\href
  {https://doi.org/10.1088/1475-7516/2016/10/001} {\bibfield  {journal}
  {\bibinfo  {journal} {JCAP}\ }\textbf {\bibinfo {volume} {10}},\ \bibinfo
  {pages} {001}},\ \Eprint {https://arxiv.org/abs/1605.01209} {arXiv:1605.01209
  [hep-ph]} \BibitemShut {NoStop}%
\bibitem [{\citenamefont {Landini}\ and\ \citenamefont
  {Strumia}(2025)}]{Landini:2025jgj}%
  \BibitemOpen
  \bibfield  {author} {\bibinfo {author} {\bibfnamefont {G.}~\bibnamefont
  {Landini}}\ and\ \bibinfo {author} {\bibfnamefont {A.}~\bibnamefont
  {Strumia}},\ }\bibfield  {title} {\bibinfo {title} {{Optical gravitational
  waves as signals of Gravitationally-Decaying Particles}},\ }\href@noop {} {\
  (\bibinfo {year} {2025})},\ \Eprint {https://arxiv.org/abs/2501.09794}
  {arXiv:2501.09794 [hep-ph]} \BibitemShut {NoStop}%
\bibitem [{\citenamefont {Fujita}\ \emph {et~al.}(2020)\citenamefont {Fujita},
  \citenamefont {Kamada},\ and\ \citenamefont {Nakai}}]{Fujita:2020rdx}%
  \BibitemOpen
  \bibfield  {author} {\bibinfo {author} {\bibfnamefont {T.}~\bibnamefont
  {Fujita}}, \bibinfo {author} {\bibfnamefont {K.}~\bibnamefont {Kamada}},\
  and\ \bibinfo {author} {\bibfnamefont {Y.}~\bibnamefont {Nakai}},\ }\bibfield
   {title} {\bibinfo {title} {{Gravitational Waves from Primordial Magnetic
  Fields via Photon-Graviton Conversion}},\ }\href
  {https://doi.org/10.1103/PhysRevD.102.103501} {\bibfield  {journal} {\bibinfo
   {journal} {Phys. Rev. D}\ }\textbf {\bibinfo {volume} {102}},\ \bibinfo
  {pages} {103501} (\bibinfo {year} {2020})},\ \Eprint
  {https://arxiv.org/abs/2002.07548} {arXiv:2002.07548 [astro-ph.CO]}
  \BibitemShut {NoStop}%
\bibitem [{\citenamefont {Servant}\ and\ \citenamefont
  {Simakachorn}(2024)}]{Servant:2023tua}%
  \BibitemOpen
  \bibfield  {author} {\bibinfo {author} {\bibfnamefont {G.}~\bibnamefont
  {Servant}}\ and\ \bibinfo {author} {\bibfnamefont {P.}~\bibnamefont
  {Simakachorn}},\ }\bibfield  {title} {\bibinfo {title} {{Ultrahigh frequency
  primordial gravitational waves beyond the kHz: The case of cosmic strings}},\
  }\href {https://doi.org/10.1103/PhysRevD.109.103538} {\bibfield  {journal}
  {\bibinfo  {journal} {Phys. Rev. D}\ }\textbf {\bibinfo {volume} {109}},\
  \bibinfo {pages} {103538} (\bibinfo {year} {2024})},\ \Eprint
  {https://arxiv.org/abs/2312.09281} {arXiv:2312.09281 [hep-ph]} \BibitemShut
  {NoStop}%
\bibitem [{\citenamefont {Delgado}\ \emph {et~al.}(2025)\citenamefont
  {Delgado}, \citenamefont {Ganz}, \citenamefont {Lin},\ and\ \citenamefont
  {Th\'eriault}}]{Delgado:2025ext}%
  \BibitemOpen
  \bibfield  {author} {\bibinfo {author} {\bibfnamefont {P.~C.~M.}\
  \bibnamefont {Delgado}}, \bibinfo {author} {\bibfnamefont {A.}~\bibnamefont
  {Ganz}}, \bibinfo {author} {\bibfnamefont {C.}~\bibnamefont {Lin}},\ and\
  \bibinfo {author} {\bibfnamefont {R.}~\bibnamefont {Th\'eriault}},\
  }\bibfield  {title} {\bibinfo {title} {{Constraining the Gravitational Wave
  Speed in the Early Universe via Gravitational Cherenkov Radiation}},\
  }\href@noop {} {\  (\bibinfo {year} {2025})},\ \Eprint
  {https://arxiv.org/abs/2501.01910} {arXiv:2501.01910 [gr-qc]} \BibitemShut
  {NoStop}%
\bibitem [{\citenamefont {Bernal}\ \emph {et~al.}(2024)\citenamefont {Bernal},
  \citenamefont {Cl\'ery}, \citenamefont {Mambrini},\ and\ \citenamefont
  {Xu}}]{Bernal:2023wus}%
  \BibitemOpen
  \bibfield  {author} {\bibinfo {author} {\bibfnamefont {N.}~\bibnamefont
  {Bernal}}, \bibinfo {author} {\bibfnamefont {S.}~\bibnamefont {Cl\'ery}},
  \bibinfo {author} {\bibfnamefont {Y.}~\bibnamefont {Mambrini}},\ and\
  \bibinfo {author} {\bibfnamefont {Y.}~\bibnamefont {Xu}},\ }\bibfield
  {title} {\bibinfo {title} {{Probing reheating with graviton
  bremsstrahlung}},\ }\href {https://doi.org/10.1088/1475-7516/2024/01/065}
  {\bibfield  {journal} {\bibinfo  {journal} {JCAP}\ }\textbf {\bibinfo
  {volume} {01}},\ \bibinfo {pages} {065}},\ \Eprint
  {https://arxiv.org/abs/2311.12694} {arXiv:2311.12694 [hep-ph]} \BibitemShut
  {NoStop}%
\bibitem [{\citenamefont {Garcia}\ and\ \citenamefont
  {Pierre}(2024)}]{Garcia:2024zir}%
  \BibitemOpen
  \bibfield  {author} {\bibinfo {author} {\bibfnamefont {M.~A.~G.}\
  \bibnamefont {Garcia}}\ and\ \bibinfo {author} {\bibfnamefont
  {M.}~\bibnamefont {Pierre}},\ }\bibfield  {title} {\bibinfo {title}
  {{Gravitational wave signatures of post-fragmentation reheating}},\ }\href
  {https://doi.org/10.1088/1475-7516/2024/09/054} {\bibfield  {journal}
  {\bibinfo  {journal} {JCAP}\ }\textbf {\bibinfo {volume} {09}},\ \bibinfo
  {pages} {054}},\ \Eprint {https://arxiv.org/abs/2404.16932} {arXiv:2404.16932
  [hep-ph]} \BibitemShut {NoStop}%
\bibitem [{\citenamefont {Saha}\ and\ \citenamefont
  {Urakawa}(2024)}]{Saha:2024lil}%
  \BibitemOpen
  \bibfield  {author} {\bibinfo {author} {\bibfnamefont {P.}~\bibnamefont
  {Saha}}\ and\ \bibinfo {author} {\bibfnamefont {Y.}~\bibnamefont {Urakawa}},\
  }\bibfield  {title} {\bibinfo {title} {{Potential Surge Preheating: enhanced
  resonance from potential features}},\ }\href@noop {} {\  (\bibinfo {year}
  {2024})},\ \Eprint {https://arxiv.org/abs/2412.17359} {arXiv:2412.17359
  [astro-ph.CO]} \BibitemShut {NoStop}%
\bibitem [{\citenamefont {Arvanitaki}\ \emph {et~al.}(2010)\citenamefont
  {Arvanitaki}, \citenamefont {Dimopoulos}, \citenamefont {Dubovsky},
  \citenamefont {Kaloper},\ and\ \citenamefont
  {March-Russell}}]{Arvanitaki:2009fg}%
  \BibitemOpen
  \bibfield  {author} {\bibinfo {author} {\bibfnamefont {A.}~\bibnamefont
  {Arvanitaki}}, \bibinfo {author} {\bibfnamefont {S.}~\bibnamefont
  {Dimopoulos}}, \bibinfo {author} {\bibfnamefont {S.}~\bibnamefont
  {Dubovsky}}, \bibinfo {author} {\bibfnamefont {N.}~\bibnamefont {Kaloper}},\
  and\ \bibinfo {author} {\bibfnamefont {J.}~\bibnamefont {March-Russell}},\
  }\bibfield  {title} {\bibinfo {title} {{String Axiverse}},\ }\href
  {https://doi.org/10.1103/PhysRevD.81.123530} {\bibfield  {journal} {\bibinfo
  {journal} {Phys. Rev. D}\ }\textbf {\bibinfo {volume} {81}},\ \bibinfo
  {pages} {123530} (\bibinfo {year} {2010})},\ \Eprint
  {https://arxiv.org/abs/0905.4720} {arXiv:0905.4720 [hep-th]} \BibitemShut
  {NoStop}%
\bibitem [{\citenamefont {Arvanitaki}\ and\ \citenamefont
  {Dubovsky}(2011)}]{Arvanitaki:2010sy}%
  \BibitemOpen
  \bibfield  {author} {\bibinfo {author} {\bibfnamefont {A.}~\bibnamefont
  {Arvanitaki}}\ and\ \bibinfo {author} {\bibfnamefont {S.}~\bibnamefont
  {Dubovsky}},\ }\bibfield  {title} {\bibinfo {title} {{Exploring the String
  Axiverse with Precision Black Hole Physics}},\ }\href
  {https://doi.org/10.1103/PhysRevD.83.044026} {\bibfield  {journal} {\bibinfo
  {journal} {Phys. Rev. D}\ }\textbf {\bibinfo {volume} {83}},\ \bibinfo
  {pages} {044026} (\bibinfo {year} {2011})},\ \Eprint
  {https://arxiv.org/abs/1004.3558} {arXiv:1004.3558 [hep-th]} \BibitemShut
  {NoStop}%
\bibitem [{\citenamefont {Yoshino}\ and\ \citenamefont
  {Kodama}(2014)}]{Yoshino:2013ofa}%
  \BibitemOpen
  \bibfield  {author} {\bibinfo {author} {\bibfnamefont {H.}~\bibnamefont
  {Yoshino}}\ and\ \bibinfo {author} {\bibfnamefont {H.}~\bibnamefont
  {Kodama}},\ }\bibfield  {title} {\bibinfo {title} {{Gravitational radiation
  from an axion cloud around a black hole: Superradiant phase}},\ }\href
  {https://doi.org/10.1093/ptep/ptu029} {\bibfield  {journal} {\bibinfo
  {journal} {PTEP}\ }\textbf {\bibinfo {volume} {2014}},\ \bibinfo {pages}
  {043E02} (\bibinfo {year} {2014})},\ \Eprint
  {https://arxiv.org/abs/1312.2326} {arXiv:1312.2326 [gr-qc]} \BibitemShut
  {NoStop}%
\bibitem [{\citenamefont {Arvanitaki}\ and\ \citenamefont
  {Geraci}(2013)}]{Arvanitaki:2012cn}%
  \BibitemOpen
  \bibfield  {author} {\bibinfo {author} {\bibfnamefont {A.}~\bibnamefont
  {Arvanitaki}}\ and\ \bibinfo {author} {\bibfnamefont {A.~A.}\ \bibnamefont
  {Geraci}},\ }\bibfield  {title} {\bibinfo {title} {{Detecting high-frequency
  gravitational waves with optically-levitated sensors}},\ }\href
  {https://doi.org/10.1103/PhysRevLett.110.071105} {\bibfield  {journal}
  {\bibinfo  {journal} {Phys. Rev. Lett.}\ }\textbf {\bibinfo {volume} {110}},\
  \bibinfo {pages} {071105} (\bibinfo {year} {2013})},\ \Eprint
  {https://arxiv.org/abs/1207.5320} {arXiv:1207.5320 [gr-qc]} \BibitemShut
  {NoStop}%
\bibitem [{\citenamefont {Aggarwal}\ \emph {et~al.}(2022)\citenamefont
  {Aggarwal}, \citenamefont {Winstone}, \citenamefont {Teo}, \citenamefont
  {Baryakhtar}, \citenamefont {Larson}, \citenamefont {Kalogera},\ and\
  \citenamefont {Geraci}}]{Aggarwal:2020umq}%
  \BibitemOpen
  \bibfield  {author} {\bibinfo {author} {\bibfnamefont {N.}~\bibnamefont
  {Aggarwal}}, \bibinfo {author} {\bibfnamefont {G.~P.}\ \bibnamefont
  {Winstone}}, \bibinfo {author} {\bibfnamefont {M.}~\bibnamefont {Teo}},
  \bibinfo {author} {\bibfnamefont {M.}~\bibnamefont {Baryakhtar}}, \bibinfo
  {author} {\bibfnamefont {S.~L.}\ \bibnamefont {Larson}}, \bibinfo {author}
  {\bibfnamefont {V.}~\bibnamefont {Kalogera}},\ and\ \bibinfo {author}
  {\bibfnamefont {A.~A.}\ \bibnamefont {Geraci}},\ }\bibfield  {title}
  {\bibinfo {title} {{Searching for New Physics with a Levitated-Sensor-Based
  Gravitational-Wave Detector}},\ }\href
  {https://doi.org/10.1103/PhysRevLett.128.111101} {\bibfield  {journal}
  {\bibinfo  {journal} {Phys. Rev. Lett.}\ }\textbf {\bibinfo {volume} {128}},\
  \bibinfo {pages} {111101} (\bibinfo {year} {2022})},\ \Eprint
  {https://arxiv.org/abs/2010.13157} {arXiv:2010.13157 [gr-qc]} \BibitemShut
  {NoStop}%
\bibitem [{\citenamefont {Carney}\ \emph {et~al.}(2024)\citenamefont {Carney},
  \citenamefont {Higgins}, \citenamefont {Marocco},\ and\ \citenamefont
  {Wentzel}}]{Carney:2024zzk}%
  \BibitemOpen
  \bibfield  {author} {\bibinfo {author} {\bibfnamefont {D.}~\bibnamefont
  {Carney}}, \bibinfo {author} {\bibfnamefont {G.}~\bibnamefont {Higgins}},
  \bibinfo {author} {\bibfnamefont {G.}~\bibnamefont {Marocco}},\ and\ \bibinfo
  {author} {\bibfnamefont {M.}~\bibnamefont {Wentzel}},\ }\bibfield  {title}
  {\bibinfo {title} {{A Superconducting Levitated Detector of Gravitational
  Waves}},\ }\href@noop {} {\  (\bibinfo {year} {2024})},\ \Eprint
  {https://arxiv.org/abs/2408.01583} {arXiv:2408.01583 [hep-ph]} \BibitemShut
  {NoStop}%
\bibitem [{\citenamefont {Goryachev}\ and\ \citenamefont
  {Tobar}(2014)}]{Goryachev:2014yra}%
  \BibitemOpen
  \bibfield  {author} {\bibinfo {author} {\bibfnamefont {M.}~\bibnamefont
  {Goryachev}}\ and\ \bibinfo {author} {\bibfnamefont {M.~E.}\ \bibnamefont
  {Tobar}},\ }\bibfield  {title} {\bibinfo {title} {{Gravitational Wave
  Detection with High Frequency Phonon Trapping Acoustic Cavities}},\ }\href
  {https://doi.org/10.1103/PhysRevD.90.102005} {\bibfield  {journal} {\bibinfo
  {journal} {Phys. Rev. D}\ }\textbf {\bibinfo {volume} {90}},\ \bibinfo
  {pages} {102005} (\bibinfo {year} {2014})},\ \bibinfo {note} {[Erratum:
  Phys.Rev.D 108, 129901 (2023)]},\ \Eprint {https://arxiv.org/abs/1410.2334}
  {arXiv:1410.2334 [gr-qc]} \BibitemShut {NoStop}%
\bibitem [{\citenamefont {Kahn}\ \emph {et~al.}(2024)\citenamefont {Kahn},
  \citenamefont {Sch\"utte-Engel},\ and\ \citenamefont
  {Trickle}}]{Kahn:2023mrj}%
  \BibitemOpen
  \bibfield  {author} {\bibinfo {author} {\bibfnamefont {Y.}~\bibnamefont
  {Kahn}}, \bibinfo {author} {\bibfnamefont {J.}~\bibnamefont
  {Sch\"utte-Engel}},\ and\ \bibinfo {author} {\bibfnamefont {T.}~\bibnamefont
  {Trickle}},\ }\bibfield  {title} {\bibinfo {title} {{Searching for
  high-frequency gravitational waves with phonons}},\ }\href
  {https://doi.org/10.1103/PhysRevD.109.096023} {\bibfield  {journal} {\bibinfo
   {journal} {Phys. Rev. D}\ }\textbf {\bibinfo {volume} {109}},\ \bibinfo
  {pages} {096023} (\bibinfo {year} {2024})},\ \Eprint
  {https://arxiv.org/abs/2311.17147} {arXiv:2311.17147 [hep-ph]} \BibitemShut
  {NoStop}%
\bibitem [{\citenamefont {Ballantini}\ \emph {et~al.}(2005)\citenamefont
  {Ballantini} \emph {et~al.}}]{Ballantini:2005am}%
  \BibitemOpen
  \bibfield  {author} {\bibinfo {author} {\bibfnamefont {R.}~\bibnamefont
  {Ballantini}} \emph {et~al.},\ }\bibfield  {title} {\bibinfo {title}
  {{Microwave apparatus for gravitational waves observation}},\ }\href@noop {}
  {\  (\bibinfo {year} {2005})},\ \Eprint {https://arxiv.org/abs/gr-qc/0502054}
  {arXiv:gr-qc/0502054} \BibitemShut {NoStop}%
\bibitem [{\citenamefont {Berlin}\ \emph {et~al.}(2023)\citenamefont {Berlin},
  \citenamefont {Blas}, \citenamefont {Tito~D'Agnolo}, \citenamefont {Ellis},
  \citenamefont {Harnik}, \citenamefont {Kahn}, \citenamefont
  {Sch\"utte-Engel},\ and\ \citenamefont {Wentzel}}]{Berlin:2023grv}%
  \BibitemOpen
  \bibfield  {author} {\bibinfo {author} {\bibfnamefont {A.}~\bibnamefont
  {Berlin}}, \bibinfo {author} {\bibfnamefont {D.}~\bibnamefont {Blas}},
  \bibinfo {author} {\bibfnamefont {R.}~\bibnamefont {Tito~D'Agnolo}}, \bibinfo
  {author} {\bibfnamefont {S.~A.~R.}\ \bibnamefont {Ellis}}, \bibinfo {author}
  {\bibfnamefont {R.}~\bibnamefont {Harnik}}, \bibinfo {author} {\bibfnamefont
  {Y.}~\bibnamefont {Kahn}}, \bibinfo {author} {\bibfnamefont {J.}~\bibnamefont
  {Sch\"utte-Engel}},\ and\ \bibinfo {author} {\bibfnamefont {M.}~\bibnamefont
  {Wentzel}},\ }\bibfield  {title} {\bibinfo {title} {{Electromagnetic cavities
  as mechanical bars for gravitational waves}},\ }\href
  {https://doi.org/10.1103/PhysRevD.108.084058} {\bibfield  {journal} {\bibinfo
   {journal} {Phys. Rev. D}\ }\textbf {\bibinfo {volume} {108}},\ \bibinfo
  {pages} {084058} (\bibinfo {year} {2023})},\ \Eprint
  {https://arxiv.org/abs/2303.01518} {arXiv:2303.01518 [hep-ph]} \BibitemShut
  {NoStop}%
\bibitem [{\citenamefont {Domcke}\ \emph
  {et~al.}(2024{\natexlab{a}})\citenamefont {Domcke}, \citenamefont {Ellis},\
  and\ \citenamefont {Rodd}}]{Domcke:2024mfu}%
  \BibitemOpen
  \bibfield  {author} {\bibinfo {author} {\bibfnamefont {V.}~\bibnamefont
  {Domcke}}, \bibinfo {author} {\bibfnamefont {S.~A.~R.}\ \bibnamefont
  {Ellis}},\ and\ \bibinfo {author} {\bibfnamefont {N.~L.}\ \bibnamefont
  {Rodd}},\ }\bibfield  {title} {\bibinfo {title} {{Magnets are Weber Bar
  Gravitational Wave Detectors}},\ }\href@noop {} {\  (\bibinfo {year}
  {2024}{\natexlab{a}})},\ \Eprint {https://arxiv.org/abs/2408.01483}
  {arXiv:2408.01483 [hep-ph]} \BibitemShut {NoStop}%
\bibitem [{\citenamefont {Schnabel}\ and\ \citenamefont
  {Korobko}(2024)}]{Schnabel:2024hem}%
  \BibitemOpen
  \bibfield  {author} {\bibinfo {author} {\bibfnamefont {R.}~\bibnamefont
  {Schnabel}}\ and\ \bibinfo {author} {\bibfnamefont {M.}~\bibnamefont
  {Korobko}},\ }\bibfield  {title} {\bibinfo {title} {{Optical sensitivities of
  current gravitational wave observatories at higher kHz, MHz and GHz
  frequencies}},\ }\href@noop {} {\  (\bibinfo {year} {2024})},\ \Eprint
  {https://arxiv.org/abs/2409.03019} {arXiv:2409.03019 [astro-ph.IM]}
  \BibitemShut {NoStop}%
\bibitem [{\citenamefont {Quach}(2016)}]{Quach:2016uxd}%
  \BibitemOpen
  \bibfield  {author} {\bibinfo {author} {\bibfnamefont {J.~Q.}\ \bibnamefont
  {Quach}},\ }\bibfield  {title} {\bibinfo {title} {{Spin gravitational
  resonance and graviton detection}},\ }\href
  {https://doi.org/10.1103/PhysRevD.93.104048} {\bibfield  {journal} {\bibinfo
  {journal} {Phys. Rev. D}\ }\textbf {\bibinfo {volume} {93}},\ \bibinfo
  {pages} {104048} (\bibinfo {year} {2016})},\ \Eprint
  {https://arxiv.org/abs/1605.08316} {arXiv:1605.08316 [gr-qc]} \BibitemShut
  {NoStop}%
\bibitem [{\citenamefont {Ito}\ \emph {et~al.}(2020)\citenamefont {Ito},
  \citenamefont {Ikeda}, \citenamefont {Miuchi},\ and\ \citenamefont
  {Soda}}]{Ito:2019wcb}%
  \BibitemOpen
  \bibfield  {author} {\bibinfo {author} {\bibfnamefont {A.}~\bibnamefont
  {Ito}}, \bibinfo {author} {\bibfnamefont {T.}~\bibnamefont {Ikeda}}, \bibinfo
  {author} {\bibfnamefont {K.}~\bibnamefont {Miuchi}},\ and\ \bibinfo {author}
  {\bibfnamefont {J.}~\bibnamefont {Soda}},\ }\bibfield  {title} {\bibinfo
  {title} {{Probing GHz gravitational waves with graviton\textendash{}magnon
  resonance}},\ }\href {https://doi.org/10.1140/epjc/s10052-020-7735-y}
  {\bibfield  {journal} {\bibinfo  {journal} {Eur. Phys. J. C}\ }\textbf
  {\bibinfo {volume} {80}},\ \bibinfo {pages} {179} (\bibinfo {year} {2020})},\
  \Eprint {https://arxiv.org/abs/1903.04843} {arXiv:1903.04843 [gr-qc]}
  \BibitemShut {NoStop}%
\bibitem [{\citenamefont {Gertsenshtein}(1961)}]{Gertsenshtein:1961xxx}%
  \BibitemOpen
  \bibfield  {author} {\bibinfo {author} {\bibfnamefont {M.~E.}\ \bibnamefont
  {Gertsenshtein}},\ }\bibfield  {title} {\bibinfo {title} {{Wave resonance of
  light and gravitational waves}},\ }\href@noop {} {\bibfield  {journal}
  {\bibinfo  {journal} {Zh. Eksp. Theor. Fiz.}\ }\textbf {\bibinfo {volume}
  {41}},\ \bibinfo {pages} {113} (\bibinfo {year} {1961})}\BibitemShut
  {NoStop}%
\bibitem [{\citenamefont {Sikivie}(1983)}]{Sikivie:1983ip}%
  \BibitemOpen
  \bibfield  {author} {\bibinfo {author} {\bibfnamefont {P.}~\bibnamefont
  {Sikivie}},\ }\bibfield  {title} {\bibinfo {title} {{Experimental Tests of
  the Invisible Axion}},\ }\href {https://doi.org/10.1103/PhysRevLett.51.1415}
  {\bibfield  {journal} {\bibinfo  {journal} {Phys. Rev. Lett.}\ }\textbf
  {\bibinfo {volume} {51}},\ \bibinfo {pages} {1415} (\bibinfo {year}
  {1983})},\ \bibinfo {note} {[Erratum: Phys.Rev.Lett. 52, 695
  (1984)]}\BibitemShut {NoStop}%
\bibitem [{\citenamefont {Domcke}\ \emph {et~al.}(2022)\citenamefont {Domcke},
  \citenamefont {Garcia-Cely},\ and\ \citenamefont {Rodd}}]{Domcke:2022rgu}%
  \BibitemOpen
  \bibfield  {author} {\bibinfo {author} {\bibfnamefont {V.}~\bibnamefont
  {Domcke}}, \bibinfo {author} {\bibfnamefont {C.}~\bibnamefont
  {Garcia-Cely}},\ and\ \bibinfo {author} {\bibfnamefont {N.~L.}\ \bibnamefont
  {Rodd}},\ }\bibfield  {title} {\bibinfo {title} {{Novel Search for
  High-Frequency Gravitational Waves with Low-Mass Axion Haloscopes}},\ }\href
  {https://doi.org/10.1103/PhysRevLett.129.041101} {\bibfield  {journal}
  {\bibinfo  {journal} {Phys. Rev. Lett.}\ }\textbf {\bibinfo {volume} {129}},\
  \bibinfo {pages} {041101} (\bibinfo {year} {2022})},\ \Eprint
  {https://arxiv.org/abs/2202.00695} {arXiv:2202.00695 [hep-ph]} \BibitemShut
  {NoStop}%
\bibitem [{\citenamefont {Bringmann}\ \emph {et~al.}(2023)\citenamefont
  {Bringmann}, \citenamefont {Domcke}, \citenamefont {Fuchs},\ and\
  \citenamefont {Kopp}}]{Bringmann:2023gba}%
  \BibitemOpen
  \bibfield  {author} {\bibinfo {author} {\bibfnamefont {T.}~\bibnamefont
  {Bringmann}}, \bibinfo {author} {\bibfnamefont {V.}~\bibnamefont {Domcke}},
  \bibinfo {author} {\bibfnamefont {E.}~\bibnamefont {Fuchs}},\ and\ \bibinfo
  {author} {\bibfnamefont {J.}~\bibnamefont {Kopp}},\ }\bibfield  {title}
  {\bibinfo {title} {{High-frequency gravitational wave detection via optical
  frequency modulation}},\ }\href
  {https://doi.org/10.1103/PhysRevD.108.L061303} {\bibfield  {journal}
  {\bibinfo  {journal} {Phys. Rev. D}\ }\textbf {\bibinfo {volume} {108}},\
  \bibinfo {pages} {L061303} (\bibinfo {year} {2023})},\ \Eprint
  {https://arxiv.org/abs/2304.10579} {arXiv:2304.10579 [hep-ph]} \BibitemShut
  {NoStop}%
\bibitem [{\citenamefont {Domcke}\ \emph
  {et~al.}(2024{\natexlab{b}})\citenamefont {Domcke}, \citenamefont
  {Garcia-Cely}, \citenamefont {Lee},\ and\ \citenamefont
  {Rodd}}]{Domcke:2023bat}%
  \BibitemOpen
  \bibfield  {author} {\bibinfo {author} {\bibfnamefont {V.}~\bibnamefont
  {Domcke}}, \bibinfo {author} {\bibfnamefont {C.}~\bibnamefont {Garcia-Cely}},
  \bibinfo {author} {\bibfnamefont {S.~M.}\ \bibnamefont {Lee}},\ and\ \bibinfo
  {author} {\bibfnamefont {N.~L.}\ \bibnamefont {Rodd}},\ }\bibfield  {title}
  {\bibinfo {title} {{Symmetries and selection rules: optimising axion
  haloscopes for Gravitational Wave searches}},\ }\href
  {https://doi.org/10.1007/JHEP03(2024)128} {\bibfield  {journal} {\bibinfo
  {journal} {JHEP}\ }\textbf {\bibinfo {volume} {03}},\ \bibinfo {pages}
  {128}},\ \Eprint {https://arxiv.org/abs/2306.03125} {arXiv:2306.03125
  [hep-ph]} \BibitemShut {NoStop}%
\bibitem [{\citenamefont {Berlin}\ \emph {et~al.}(2022)\citenamefont {Berlin},
  \citenamefont {Blas}, \citenamefont {Tito~D'Agnolo}, \citenamefont {Ellis},
  \citenamefont {Harnik}, \citenamefont {Kahn},\ and\ \citenamefont
  {Sch\"utte-Engel}}]{Berlin:2021txa}%
  \BibitemOpen
  \bibfield  {author} {\bibinfo {author} {\bibfnamefont {A.}~\bibnamefont
  {Berlin}}, \bibinfo {author} {\bibfnamefont {D.}~\bibnamefont {Blas}},
  \bibinfo {author} {\bibfnamefont {R.}~\bibnamefont {Tito~D'Agnolo}}, \bibinfo
  {author} {\bibfnamefont {S.~A.~R.}\ \bibnamefont {Ellis}}, \bibinfo {author}
  {\bibfnamefont {R.}~\bibnamefont {Harnik}}, \bibinfo {author} {\bibfnamefont
  {Y.}~\bibnamefont {Kahn}},\ and\ \bibinfo {author} {\bibfnamefont
  {J.}~\bibnamefont {Sch\"utte-Engel}},\ }\bibfield  {title} {\bibinfo {title}
  {{Detecting high-frequency gravitational waves with microwave cavities}},\
  }\href {https://doi.org/10.1103/PhysRevD.105.116011} {\bibfield  {journal}
  {\bibinfo  {journal} {Phys. Rev. D}\ }\textbf {\bibinfo {volume} {105}},\
  \bibinfo {pages} {116011} (\bibinfo {year} {2022})},\ \Eprint
  {https://arxiv.org/abs/2112.11465} {arXiv:2112.11465 [hep-ph]} \BibitemShut
  {NoStop}%
\bibitem [{\citenamefont {Navarro}\ \emph {et~al.}(2024)\citenamefont
  {Navarro}, \citenamefont {Gimeno}, \citenamefont {Monz\'o-Cabrera},
  \citenamefont {D\'\i{}az-Morcillo},\ and\ \citenamefont
  {Blas}}]{Navarro:2023eii}%
  \BibitemOpen
  \bibfield  {author} {\bibinfo {author} {\bibfnamefont {P.}~\bibnamefont
  {Navarro}}, \bibinfo {author} {\bibfnamefont {B.}~\bibnamefont {Gimeno}},
  \bibinfo {author} {\bibfnamefont {J.}~\bibnamefont {Monz\'o-Cabrera}},
  \bibinfo {author} {\bibfnamefont {A.}~\bibnamefont {D\'\i{}az-Morcillo}},\
  and\ \bibinfo {author} {\bibfnamefont {D.}~\bibnamefont {Blas}},\ }\bibfield
  {title} {\bibinfo {title} {{Study of a cubic cavity resonator for
  gravitational waves detection in the microwave frequency range}},\ }\href
  {https://doi.org/10.1103/PhysRevD.109.104048} {\bibfield  {journal} {\bibinfo
   {journal} {Phys. Rev. D}\ }\textbf {\bibinfo {volume} {109}},\ \bibinfo
  {pages} {104048} (\bibinfo {year} {2024})},\ \Eprint
  {https://arxiv.org/abs/2312.02270} {arXiv:2312.02270 [hep-ph]} \BibitemShut
  {NoStop}%
\bibitem [{\citenamefont {Ahn}\ \emph {et~al.}(2024)\citenamefont {Ahn},
  \citenamefont {Bae}, \citenamefont {Im},\ and\ \citenamefont
  {Park}}]{Ahn:2023mrg}%
  \BibitemOpen
  \bibfield  {author} {\bibinfo {author} {\bibfnamefont {D.}~\bibnamefont
  {Ahn}}, \bibinfo {author} {\bibfnamefont {Y.-B.}\ \bibnamefont {Bae}},
  \bibinfo {author} {\bibfnamefont {S.~H.}\ \bibnamefont {Im}},\ and\ \bibinfo
  {author} {\bibfnamefont {C.}~\bibnamefont {Park}},\ }\bibfield  {title}
  {\bibinfo {title} {{Electromagnetic field in a cavity induced by
  gravitational waves}},\ }\href {https://doi.org/10.1103/PhysRevD.110.064061}
  {\bibfield  {journal} {\bibinfo  {journal} {Phys. Rev. D}\ }\textbf {\bibinfo
  {volume} {110}},\ \bibinfo {pages} {064061} (\bibinfo {year} {2024})},\
  \Eprint {https://arxiv.org/abs/2312.09550} {arXiv:2312.09550 [gr-qc]}
  \BibitemShut {NoStop}%
\bibitem [{\citenamefont {Gatti}\ \emph {et~al.}(2024)\citenamefont {Gatti},
  \citenamefont {Visinelli},\ and\ \citenamefont
  {Zantedeschi}}]{Gatti:2024mde}%
  \BibitemOpen
  \bibfield  {author} {\bibinfo {author} {\bibfnamefont {C.}~\bibnamefont
  {Gatti}}, \bibinfo {author} {\bibfnamefont {L.}~\bibnamefont {Visinelli}},\
  and\ \bibinfo {author} {\bibfnamefont {M.}~\bibnamefont {Zantedeschi}},\
  }\bibfield  {title} {\bibinfo {title} {{Cavity detection of gravitational
  waves: Where do we stand?}},\ }\href
  {https://doi.org/10.1103/PhysRevD.110.023018} {\bibfield  {journal} {\bibinfo
   {journal} {Phys. Rev. D}\ }\textbf {\bibinfo {volume} {110}},\ \bibinfo
  {pages} {023018} (\bibinfo {year} {2024})},\ \Eprint
  {https://arxiv.org/abs/2403.18610} {arXiv:2403.18610 [gr-qc]} \BibitemShut
  {NoStop}%
\bibitem [{\citenamefont {Domcke}\ \emph
  {et~al.}(2024{\natexlab{c}})\citenamefont {Domcke}, \citenamefont {Ellis},\
  and\ \citenamefont {Kopp}}]{Domcke:2024eti}%
  \BibitemOpen
  \bibfield  {author} {\bibinfo {author} {\bibfnamefont {V.}~\bibnamefont
  {Domcke}}, \bibinfo {author} {\bibfnamefont {S.~A.~R.}\ \bibnamefont
  {Ellis}},\ and\ \bibinfo {author} {\bibfnamefont {J.}~\bibnamefont {Kopp}},\
  }\bibfield  {title} {\bibinfo {title} {{Dielectric Haloscopes as
  Gravitational Wave Detectors}},\ }\href@noop {} {\  (\bibinfo {year}
  {2024}{\natexlab{c}})},\ \Eprint {https://arxiv.org/abs/2409.06462}
  {arXiv:2409.06462 [hep-ph]} \BibitemShut {NoStop}%
\bibitem [{\citenamefont {Capdevilla}\ \emph {et~al.}(2024)\citenamefont
  {Capdevilla}, \citenamefont {Gelmini}, \citenamefont {Hyman}, \citenamefont
  {Millar},\ and\ \citenamefont {Vitagliano}}]{Capdevilla:2024cby}%
  \BibitemOpen
  \bibfield  {author} {\bibinfo {author} {\bibfnamefont {R.}~\bibnamefont
  {Capdevilla}}, \bibinfo {author} {\bibfnamefont {G.~B.}\ \bibnamefont
  {Gelmini}}, \bibinfo {author} {\bibfnamefont {J.}~\bibnamefont {Hyman}},
  \bibinfo {author} {\bibfnamefont {A.~J.}\ \bibnamefont {Millar}},\ and\
  \bibinfo {author} {\bibfnamefont {E.}~\bibnamefont {Vitagliano}},\ }\bibfield
   {title} {\bibinfo {title} {{Gravitational Wave Detection With Plasma
  Haloscopes}},\ }\href@noop {} {\  (\bibinfo {year} {2024})},\ \Eprint
  {https://arxiv.org/abs/2412.14450} {arXiv:2412.14450 [hep-ph]} \BibitemShut
  {NoStop}%
\bibitem [{\citenamefont {Ejlli}\ \emph {et~al.}(2019)\citenamefont {Ejlli},
  \citenamefont {Ejlli}, \citenamefont {Cruise}, \citenamefont {Pisano},\ and\
  \citenamefont {Grote}}]{Ejlli:2019bqj}%
  \BibitemOpen
  \bibfield  {author} {\bibinfo {author} {\bibfnamefont {A.}~\bibnamefont
  {Ejlli}}, \bibinfo {author} {\bibfnamefont {D.}~\bibnamefont {Ejlli}},
  \bibinfo {author} {\bibfnamefont {A.~M.}\ \bibnamefont {Cruise}}, \bibinfo
  {author} {\bibfnamefont {G.}~\bibnamefont {Pisano}},\ and\ \bibinfo {author}
  {\bibfnamefont {H.}~\bibnamefont {Grote}},\ }\bibfield  {title} {\bibinfo
  {title} {{Upper limits on the amplitude of ultra-high-frequency gravitational
  waves from graviton to photon conversion}},\ }\href
  {https://doi.org/10.1140/epjc/s10052-019-7542-5} {\bibfield  {journal}
  {\bibinfo  {journal} {Eur. Phys. J. C}\ }\textbf {\bibinfo {volume} {79}},\
  \bibinfo {pages} {1032} (\bibinfo {year} {2019})},\ \Eprint
  {https://arxiv.org/abs/1908.00232} {arXiv:1908.00232 [gr-qc]} \BibitemShut
  {NoStop}%
\bibitem [{\citenamefont {Liu}\ \emph {et~al.}(2022)\citenamefont {Liu} \emph
  {et~al.}}]{BREAD:2021tpx}%
  \BibitemOpen
  \bibfield  {author} {\bibinfo {author} {\bibfnamefont {J.}~\bibnamefont
  {Liu}} \emph {et~al.} (\bibinfo {collaboration} {BREAD}),\ }\bibfield
  {title} {\bibinfo {title} {{Broadband Solenoidal Haloscope for Terahertz
  Axion Detection}},\ }\href {https://doi.org/10.1103/PhysRevLett.128.131801}
  {\bibfield  {journal} {\bibinfo  {journal} {Phys. Rev. Lett.}\ }\textbf
  {\bibinfo {volume} {128}},\ \bibinfo {pages} {131801} (\bibinfo {year}
  {2022})},\ \Eprint {https://arxiv.org/abs/2111.12103} {arXiv:2111.12103
  [physics.ins-det]} \BibitemShut {NoStop}%
\bibitem [{\citenamefont {Horns}\ \emph {et~al.}(2013)\citenamefont {Horns},
  \citenamefont {Jaeckel}, \citenamefont {Lindner}, \citenamefont {Lobanov},
  \citenamefont {Redondo},\ and\ \citenamefont {Ringwald}}]{Horns:2012jf}%
  \BibitemOpen
  \bibfield  {author} {\bibinfo {author} {\bibfnamefont {D.}~\bibnamefont
  {Horns}}, \bibinfo {author} {\bibfnamefont {J.}~\bibnamefont {Jaeckel}},
  \bibinfo {author} {\bibfnamefont {A.}~\bibnamefont {Lindner}}, \bibinfo
  {author} {\bibfnamefont {A.}~\bibnamefont {Lobanov}}, \bibinfo {author}
  {\bibfnamefont {J.}~\bibnamefont {Redondo}},\ and\ \bibinfo {author}
  {\bibfnamefont {A.}~\bibnamefont {Ringwald}},\ }\bibfield  {title} {\bibinfo
  {title} {{Searching for WISPy Cold Dark Matter with a Dish Antenna}},\ }\href
  {https://doi.org/10.1088/1475-7516/2013/04/016} {\bibfield  {journal}
  {\bibinfo  {journal} {JCAP}\ }\textbf {\bibinfo {volume} {04}},\ \bibinfo
  {pages} {016}},\ \Eprint {https://arxiv.org/abs/1212.2970} {arXiv:1212.2970
  [hep-ph]} \BibitemShut {NoStop}%
\bibitem [{\citenamefont {Suzuki}\ \emph {et~al.}(2015)\citenamefont {Suzuki},
  \citenamefont {Horie}, \citenamefont {Inoue},\ and\ \citenamefont
  {Minowa}}]{Suzuki:2015sza}%
  \BibitemOpen
  \bibfield  {author} {\bibinfo {author} {\bibfnamefont {J.}~\bibnamefont
  {Suzuki}}, \bibinfo {author} {\bibfnamefont {T.}~\bibnamefont {Horie}},
  \bibinfo {author} {\bibfnamefont {Y.}~\bibnamefont {Inoue}},\ and\ \bibinfo
  {author} {\bibfnamefont {M.}~\bibnamefont {Minowa}},\ }\bibfield  {title}
  {\bibinfo {title} {{Experimental Search for Hidden Photon CDM in the eV mass
  range with a Dish Antenna}},\ }\href
  {https://doi.org/10.1088/1475-7516/2015/09/042} {\bibfield  {journal}
  {\bibinfo  {journal} {JCAP}\ }\textbf {\bibinfo {volume} {09}},\ \bibinfo
  {pages} {042}},\ \Eprint {https://arxiv.org/abs/1504.00118} {arXiv:1504.00118
  [hep-ex]} \BibitemShut {NoStop}%
\bibitem [{\citenamefont {Jaeckel}\ and\ \citenamefont
  {Knirck}(2016)}]{Jaeckel:2015kea}%
  \BibitemOpen
  \bibfield  {author} {\bibinfo {author} {\bibfnamefont {J.}~\bibnamefont
  {Jaeckel}}\ and\ \bibinfo {author} {\bibfnamefont {S.}~\bibnamefont
  {Knirck}},\ }\bibfield  {title} {\bibinfo {title} {{Directional Resolution of
  Dish Antenna Experiments to Search for WISPy Dark Matter}},\ }\href
  {https://doi.org/10.1088/1475-7516/2016/01/005} {\bibfield  {journal}
  {\bibinfo  {journal} {JCAP}\ }\textbf {\bibinfo {volume} {01}},\ \bibinfo
  {pages} {005}},\ \Eprint {https://arxiv.org/abs/1509.00371} {arXiv:1509.00371
  [hep-ph]} \BibitemShut {NoStop}%
\bibitem [{\citenamefont {Knirck}\ \emph {et~al.}(2018)\citenamefont {Knirck},
  \citenamefont {Yamazaki}, \citenamefont {Okesaku}, \citenamefont {Asai},
  \citenamefont {Idehara},\ and\ \citenamefont {Inada}}]{Knirck:2018ojz}%
  \BibitemOpen
  \bibfield  {author} {\bibinfo {author} {\bibfnamefont {S.}~\bibnamefont
  {Knirck}}, \bibinfo {author} {\bibfnamefont {T.}~\bibnamefont {Yamazaki}},
  \bibinfo {author} {\bibfnamefont {Y.}~\bibnamefont {Okesaku}}, \bibinfo
  {author} {\bibfnamefont {S.}~\bibnamefont {Asai}}, \bibinfo {author}
  {\bibfnamefont {T.}~\bibnamefont {Idehara}},\ and\ \bibinfo {author}
  {\bibfnamefont {T.}~\bibnamefont {Inada}},\ }\bibfield  {title} {\bibinfo
  {title} {{First results from a hidden photon dark matter search in the meV
  sector using a plane-parabolic mirror system}},\ }\href
  {https://doi.org/10.1088/1475-7516/2018/11/031} {\bibfield  {journal}
  {\bibinfo  {journal} {JCAP}\ }\textbf {\bibinfo {volume} {11}},\ \bibinfo
  {pages} {031}},\ \Eprint {https://arxiv.org/abs/1806.05120} {arXiv:1806.05120
  [hep-ex]} \BibitemShut {NoStop}%
\bibitem [{\citenamefont {Andrianavalomahefa}\ \emph
  {et~al.}(2020)\citenamefont {Andrianavalomahefa} \emph
  {et~al.}}]{FUNKExperiment:2020ofv}%
  \BibitemOpen
  \bibfield  {author} {\bibinfo {author} {\bibfnamefont {A.}~\bibnamefont
  {Andrianavalomahefa}} \emph {et~al.} (\bibinfo {collaboration} {FUNK
  Experiment}),\ }\bibfield  {title} {\bibinfo {title} {{Limits from the Funk
  Experiment on the Mixing Strength of Hidden-Photon Dark Matter in the Visible
  and Near-Ultraviolet Wavelength Range}},\ }\href
  {https://doi.org/10.1103/PhysRevD.102.042001} {\bibfield  {journal} {\bibinfo
   {journal} {Phys. Rev. D}\ }\textbf {\bibinfo {volume} {102}},\ \bibinfo
  {pages} {042001} (\bibinfo {year} {2020})},\ \Eprint
  {https://arxiv.org/abs/2003.13144} {arXiv:2003.13144 [astro-ph.CO]}
  \BibitemShut {NoStop}%
\bibitem [{\citenamefont {Tomita}\ \emph {et~al.}(2020)\citenamefont {Tomita},
  \citenamefont {Oguri}, \citenamefont {Inoue}, \citenamefont {Minowa},
  \citenamefont {Nagasaki}, \citenamefont {Suzuki},\ and\ \citenamefont
  {Tajima}}]{Tomita:2020usq}%
  \BibitemOpen
  \bibfield  {author} {\bibinfo {author} {\bibfnamefont {N.}~\bibnamefont
  {Tomita}}, \bibinfo {author} {\bibfnamefont {S.}~\bibnamefont {Oguri}},
  \bibinfo {author} {\bibfnamefont {Y.}~\bibnamefont {Inoue}}, \bibinfo
  {author} {\bibfnamefont {M.}~\bibnamefont {Minowa}}, \bibinfo {author}
  {\bibfnamefont {T.}~\bibnamefont {Nagasaki}}, \bibinfo {author}
  {\bibfnamefont {J.}~\bibnamefont {Suzuki}},\ and\ \bibinfo {author}
  {\bibfnamefont {O.}~\bibnamefont {Tajima}},\ }\bibfield  {title} {\bibinfo
  {title} {{Search for hidden-photon cold dark matter using a K-band cryogenic
  receiver}},\ }\href {https://doi.org/10.1088/1475-7516/2020/09/012}
  {\bibfield  {journal} {\bibinfo  {journal} {JCAP}\ }\textbf {\bibinfo
  {volume} {09}},\ \bibinfo {pages} {012}},\ \Eprint
  {https://arxiv.org/abs/2006.02828} {arXiv:2006.02828 [hep-ex]} \BibitemShut
  {NoStop}%
\bibitem [{\citenamefont {Manasse}\ and\ \citenamefont
  {Misner}(1963)}]{Manasse:1963zz}%
  \BibitemOpen
  \bibfield  {author} {\bibinfo {author} {\bibfnamefont {F.~K.}\ \bibnamefont
  {Manasse}}\ and\ \bibinfo {author} {\bibfnamefont {C.~W.}\ \bibnamefont
  {Misner}},\ }\bibfield  {title} {\bibinfo {title} {{Fermi Normal Coordinates
  and Some Basic Concepts in Differential Geometry}},\ }\href
  {https://doi.org/10.1063/1.1724316} {\bibfield  {journal} {\bibinfo
  {journal} {J. Math. Phys.}\ }\textbf {\bibinfo {volume} {4}},\ \bibinfo
  {pages} {735} (\bibinfo {year} {1963})}\BibitemShut {NoStop}%
\bibitem [{\citenamefont {Ratzinger}\ \emph {et~al.}(2024)\citenamefont
  {Ratzinger}, \citenamefont {Schenk},\ and\ \citenamefont
  {Schwaller}}]{Ratzinger:2024spd}%
  \BibitemOpen
  \bibfield  {author} {\bibinfo {author} {\bibfnamefont {W.}~\bibnamefont
  {Ratzinger}}, \bibinfo {author} {\bibfnamefont {S.}~\bibnamefont {Schenk}},\
  and\ \bibinfo {author} {\bibfnamefont {P.}~\bibnamefont {Schwaller}},\
  }\bibfield  {title} {\bibinfo {title} {{A coordinate-independent formalism
  for detecting high-frequency gravitational waves}},\ }\href
  {https://doi.org/10.1007/JHEP08(2024)195} {\bibfield  {journal} {\bibinfo
  {journal} {JHEP}\ }\textbf {\bibinfo {volume} {08}},\ \bibinfo {pages}
  {195}},\ \Eprint {https://arxiv.org/abs/2404.08572} {arXiv:2404.08572
  [gr-qc]} \BibitemShut {NoStop}%
\bibitem [{\citenamefont {De~Logi}\ and\ \citenamefont
  {Mickelson}(1977)}]{PhysRevD.16.2915}%
  \BibitemOpen
  \bibfield  {author} {\bibinfo {author} {\bibfnamefont {W.~K.}\ \bibnamefont
  {De~Logi}}\ and\ \bibinfo {author} {\bibfnamefont {A.~R.}\ \bibnamefont
  {Mickelson}},\ }\bibfield  {title} {\bibinfo {title} {Electrogravitational
  conversion cross sections in static electromagnetic fields},\ }\href
  {https://doi.org/10.1103/PhysRevD.16.2915} {\bibfield  {journal} {\bibinfo
  {journal} {Phys. Rev. D}\ }\textbf {\bibinfo {volume} {16}},\ \bibinfo
  {pages} {2915} (\bibinfo {year} {1977})}\BibitemShut {NoStop}%
\bibitem [{com()}]{comsol}%
  \BibitemOpen
  \href@noop {} {\bibinfo {title} {Comsol
  multiphysics\textsuperscript{\tiny\textregistered} v. 6.3.}},\ \bibinfo
  {howpublished} {\url{www.comsol.com.}}\BibitemShut {Stop}%
\bibitem [{\citenamefont {Knirck}\ \emph {et~al.}(2024)\citenamefont {Knirck},
  \citenamefont {Hoshino}, \citenamefont {Awida}, \citenamefont {Cancelo},
  \citenamefont {Di~Federico}, \citenamefont {Knepper}, \citenamefont
  {Lapuente}, \citenamefont {Littmann}, \citenamefont {Miller}, \citenamefont
  {Mitchell}, \citenamefont {Rodriguez}, \citenamefont {Ruschman},
  \citenamefont {Sawtell}, \citenamefont {Stefanazzi}, \citenamefont
  {Sonnenschein}, \citenamefont {Teafoe}, \citenamefont {Bowring},
  \citenamefont {Carosi}, \citenamefont {Chou}, \citenamefont {Chang},
  \citenamefont {Dona}, \citenamefont {Khatiwada}, \citenamefont {Kurinsky},
  \citenamefont {Liu}, \citenamefont {Pena}, \citenamefont {Salemi},
  \citenamefont {Wang},\ and\ \citenamefont {Yu}}]{GigaBREAD_Axion_search}%
  \BibitemOpen
  \bibfield  {author} {\bibinfo {author} {\bibfnamefont {S.}~\bibnamefont
  {Knirck}}, \bibinfo {author} {\bibfnamefont {G.}~\bibnamefont {Hoshino}},
  \bibinfo {author} {\bibfnamefont {M.~H.}\ \bibnamefont {Awida}}, \bibinfo
  {author} {\bibfnamefont {G.~I.}\ \bibnamefont {Cancelo}}, \bibinfo {author}
  {\bibfnamefont {M.}~\bibnamefont {Di~Federico}}, \bibinfo {author}
  {\bibfnamefont {B.}~\bibnamefont {Knepper}}, \bibinfo {author} {\bibfnamefont
  {A.}~\bibnamefont {Lapuente}}, \bibinfo {author} {\bibfnamefont
  {M.}~\bibnamefont {Littmann}}, \bibinfo {author} {\bibfnamefont {D.~W.}\
  \bibnamefont {Miller}}, \bibinfo {author} {\bibfnamefont {D.~V.}\
  \bibnamefont {Mitchell}}, \bibinfo {author} {\bibfnamefont {D.}~\bibnamefont
  {Rodriguez}}, \bibinfo {author} {\bibfnamefont {M.~K.}\ \bibnamefont
  {Ruschman}}, \bibinfo {author} {\bibfnamefont {M.~A.}\ \bibnamefont
  {Sawtell}}, \bibinfo {author} {\bibfnamefont {L.}~\bibnamefont {Stefanazzi}},
  \bibinfo {author} {\bibfnamefont {A.}~\bibnamefont {Sonnenschein}}, \bibinfo
  {author} {\bibfnamefont {G.~W.}\ \bibnamefont {Teafoe}}, \bibinfo {author}
  {\bibfnamefont {D.}~\bibnamefont {Bowring}}, \bibinfo {author} {\bibfnamefont
  {G.}~\bibnamefont {Carosi}}, \bibinfo {author} {\bibfnamefont
  {A.}~\bibnamefont {Chou}}, \bibinfo {author} {\bibfnamefont {C.~L.}\
  \bibnamefont {Chang}}, \bibinfo {author} {\bibfnamefont {K.}~\bibnamefont
  {Dona}}, \bibinfo {author} {\bibfnamefont {R.}~\bibnamefont {Khatiwada}},
  \bibinfo {author} {\bibfnamefont {N.~A.}\ \bibnamefont {Kurinsky}}, \bibinfo
  {author} {\bibfnamefont {J.}~\bibnamefont {Liu}}, \bibinfo {author}
  {\bibfnamefont {C.}~\bibnamefont {Pena}}, \bibinfo {author} {\bibfnamefont
  {C.~P.}\ \bibnamefont {Salemi}}, \bibinfo {author} {\bibfnamefont {C.~W.}\
  \bibnamefont {Wang}},\ and\ \bibinfo {author} {\bibfnamefont
  {J.}~\bibnamefont {Yu}} (\bibinfo {collaboration} {BREAD Collaboration}),\
  }\bibfield  {title} {\bibinfo {title} {First results from a broadband search
  for dark photon dark matter in the 44 to $52\text{ }\text{
  }\mathrm{\ensuremath{\mu}}\mathrm{eV}$ range with a coaxial dish antenna},\
  }\href {https://doi.org/10.1103/PhysRevLett.132.131004} {\bibfield  {journal}
  {\bibinfo  {journal} {Phys. Rev. Lett.}\ }\textbf {\bibinfo {volume} {132}},\
  \bibinfo {pages} {131004} (\bibinfo {year} {2024})}\BibitemShut {NoStop}%
\bibitem [{\citenamefont {Armengaud}\ \emph {et~al.}(2014)\citenamefont
  {Armengaud} \emph {et~al.}}]{Armengaud:2014gea}%
  \BibitemOpen
  \bibfield  {author} {\bibinfo {author} {\bibfnamefont {E.}~\bibnamefont
  {Armengaud}} \emph {et~al.},\ }\bibfield  {title} {\bibinfo {title}
  {{Conceptual Design of the International Axion Observatory (IAXO)}},\ }\href
  {https://doi.org/10.1088/1748-0221/9/05/T05002} {\bibfield  {journal}
  {\bibinfo  {journal} {JINST}\ }\textbf {\bibinfo {volume} {9}},\ \bibinfo
  {pages} {T05002}},\ \Eprint {https://arxiv.org/abs/1401.3233}
  {arXiv:1401.3233 [physics.ins-det]} \BibitemShut {NoStop}%
\bibitem [{\citenamefont {Liu}\ \emph {et~al.}(2024)\citenamefont {Liu},
  \citenamefont {Ren},\ and\ \citenamefont {Zhang}}]{PhysRevLett.132.131402}%
  \BibitemOpen
  \bibfield  {author} {\bibinfo {author} {\bibfnamefont {T.}~\bibnamefont
  {Liu}}, \bibinfo {author} {\bibfnamefont {J.}~\bibnamefont {Ren}},\ and\
  \bibinfo {author} {\bibfnamefont {C.}~\bibnamefont {Zhang}},\ }\bibfield
  {title} {\bibinfo {title} {Limits on high-frequency gravitational waves in
  planetary magnetospheres},\ }\href
  {https://doi.org/10.1103/PhysRevLett.132.131402} {\bibfield  {journal}
  {\bibinfo  {journal} {Phys. Rev. Lett.}\ }\textbf {\bibinfo {volume} {132}},\
  \bibinfo {pages} {131402} (\bibinfo {year} {2024})}\BibitemShut {NoStop}%
\bibitem [{\citenamefont {Dicke}(1946)}]{Dicke}%
  \BibitemOpen
  \bibfield  {author} {\bibinfo {author} {\bibfnamefont {R.~H.}\ \bibnamefont
  {Dicke}},\ }\bibfield  {title} {\bibinfo {title} {The measurement of thermal
  radiation at microwave frequencies},\ }\href
  {https://doi.org/10.1063/1.1770483} {\bibfield  {journal} {\bibinfo
  {journal} {Review of Scientific Instruments}\ }\textbf {\bibinfo {volume}
  {17}},\ \bibinfo {pages} {268} (\bibinfo {year} {1946})},\ \Eprint
  {https://arxiv.org/abs/https://pubs.aip.org/aip/rsi/article-pdf/17/7/268/19120701/268\_1\_online.pdf}
  {https://pubs.aip.org/aip/rsi/article-pdf/17/7/268/19120701/268\_1\_online.pdf}
  \BibitemShut {NoStop}%
\bibitem [{\citenamefont {Verma}\ \emph {et~al.}(2021)\citenamefont {Verma}
  \emph {et~al.}}]{Verma:2020gso}%
  \BibitemOpen
  \bibfield  {author} {\bibinfo {author} {\bibfnamefont {V.~B.}\ \bibnamefont
  {Verma}} \emph {et~al.},\ }\bibfield  {title} {\bibinfo {title}
  {{Single-photon detection in the mid-infrared up to 10 \ensuremath{\mu}m
  wavelength using tungsten silicide superconducting nanowire detectors}},\
  }\href {https://doi.org/10.1063/5.0048049} {\bibfield  {journal} {\bibinfo
  {journal} {APL Photon.}\ }\textbf {\bibinfo {volume} {6}},\ \bibinfo {pages}
  {056101} (\bibinfo {year} {2021})},\ \Eprint
  {https://arxiv.org/abs/2012.09979} {arXiv:2012.09979 [physics.ins-det]}
  \BibitemShut {NoStop}%
\bibitem [{\citenamefont {Wollman}\ \emph {et~al.}(2017)\citenamefont
  {Wollman}, \citenamefont {Verma}, \citenamefont {Beyer}, \citenamefont
  {Briggs}, \citenamefont {Korzh}, \citenamefont {Allmaras}, \citenamefont
  {Marsili}, \citenamefont {Lita}, \citenamefont {Mirin}, \citenamefont {Nam},\
  and\ \citenamefont {Shaw}}]{Wollman:17}%
  \BibitemOpen
  \bibfield  {author} {\bibinfo {author} {\bibfnamefont {E.~E.}\ \bibnamefont
  {Wollman}}, \bibinfo {author} {\bibfnamefont {V.~B.}\ \bibnamefont {Verma}},
  \bibinfo {author} {\bibfnamefont {A.~D.}\ \bibnamefont {Beyer}}, \bibinfo
  {author} {\bibfnamefont {R.~M.}\ \bibnamefont {Briggs}}, \bibinfo {author}
  {\bibfnamefont {B.}~\bibnamefont {Korzh}}, \bibinfo {author} {\bibfnamefont
  {J.~P.}\ \bibnamefont {Allmaras}}, \bibinfo {author} {\bibfnamefont
  {F.}~\bibnamefont {Marsili}}, \bibinfo {author} {\bibfnamefont {A.~E.}\
  \bibnamefont {Lita}}, \bibinfo {author} {\bibfnamefont {R.~P.}\ \bibnamefont
  {Mirin}}, \bibinfo {author} {\bibfnamefont {S.~W.}\ \bibnamefont {Nam}},\
  and\ \bibinfo {author} {\bibfnamefont {M.~D.}\ \bibnamefont {Shaw}},\
  }\bibfield  {title} {\bibinfo {title} {Uv superconducting nanowire
  single-photon detectors with high efficiency, low noise, and 4 k operating
  temperature},\ }\href {https://doi.org/10.1364/OE.25.026792} {\bibfield
  {journal} {\bibinfo  {journal} {Opt. Express}\ }\textbf {\bibinfo {volume}
  {25}},\ \bibinfo {pages} {26792} (\bibinfo {year} {2017})}\BibitemShut
  {NoStop}%
\bibitem [{\citenamefont {Lamoreaux}\ \emph {et~al.}(2013)\citenamefont
  {Lamoreaux}, \citenamefont {van Bibber}, \citenamefont {Lehnert},\ and\
  \citenamefont {Carosi}}]{PhysRevD.88.035020}%
  \BibitemOpen
  \bibfield  {author} {\bibinfo {author} {\bibfnamefont {S.~K.}\ \bibnamefont
  {Lamoreaux}}, \bibinfo {author} {\bibfnamefont {K.~A.}\ \bibnamefont {van
  Bibber}}, \bibinfo {author} {\bibfnamefont {K.~W.}\ \bibnamefont {Lehnert}},\
  and\ \bibinfo {author} {\bibfnamefont {G.}~\bibnamefont {Carosi}},\
  }\bibfield  {title} {\bibinfo {title} {Analysis of single-photon and linear
  amplifier detectors for microwave cavity dark matter axion searches},\ }\href
  {https://doi.org/10.1103/PhysRevD.88.035020} {\bibfield  {journal} {\bibinfo
  {journal} {Phys. Rev. D}\ }\textbf {\bibinfo {volume} {88}},\ \bibinfo
  {pages} {035020} (\bibinfo {year} {2013})}\BibitemShut {NoStop}%
\bibitem [{\citenamefont {Braggio}\ \emph {et~al.}(2024)\citenamefont {Braggio}
  \emph {et~al.}}]{Braggio:2024xed}%
  \BibitemOpen
  \bibfield  {author} {\bibinfo {author} {\bibfnamefont {C.}~\bibnamefont
  {Braggio}} \emph {et~al.},\ }\bibfield  {title} {\bibinfo {title}
  {{Quantum-enhanced sensing of axion dark matter with a transmon-based single
  microwave photon counter}},\ }\href@noop {} {\  (\bibinfo {year} {2024})},\
  \Eprint {https://arxiv.org/abs/2403.02321} {arXiv:2403.02321 [quant-ph]}
  \BibitemShut {NoStop}%
\bibitem [{\citenamefont {B\"ahre}\ \emph {et~al.}(2013)\citenamefont {B\"ahre}
  \emph {et~al.}}]{Bahre:2013ywa}%
  \BibitemOpen
  \bibfield  {author} {\bibinfo {author} {\bibfnamefont {R.}~\bibnamefont
  {B\"ahre}} \emph {et~al.},\ }\bibfield  {title} {\bibinfo {title} {{Any light
  particle search II \textemdash{}Technical Design Report}},\ }\href
  {https://doi.org/10.1088/1748-0221/8/09/T09001} {\bibfield  {journal}
  {\bibinfo  {journal} {JINST}\ }\textbf {\bibinfo {volume} {8}},\ \bibinfo
  {pages} {T09001}},\ \Eprint {https://arxiv.org/abs/1302.5647}
  {arXiv:1302.5647 [physics.ins-det]} \BibitemShut {NoStop}%
\bibitem [{\citenamefont {Brun}\ \emph {et~al.}(2019)\citenamefont {Brun} \emph
  {et~al.}}]{MADMAX:2019pub}%
  \BibitemOpen
  \bibfield  {author} {\bibinfo {author} {\bibfnamefont {P.}~\bibnamefont
  {Brun}} \emph {et~al.} (\bibinfo {collaboration} {MADMAX}),\ }\bibfield
  {title} {\bibinfo {title} {{A new experimental approach to probe QCD axion
  dark matter in the mass range above 40 $\mu$eV}},\ }\href
  {https://doi.org/10.1140/epjc/s10052-019-6683-x} {\bibfield  {journal}
  {\bibinfo  {journal} {Eur. Phys. J. C}\ }\textbf {\bibinfo {volume} {79}},\
  \bibinfo {pages} {186} (\bibinfo {year} {2019})},\ \Eprint
  {https://arxiv.org/abs/1901.07401} {arXiv:1901.07401 [physics.ins-det]}
  \BibitemShut {NoStop}%
\bibitem [{\citenamefont {Egge}(2023)}]{Egge_2023}%
  \BibitemOpen
  \bibfield  {author} {\bibinfo {author} {\bibfnamefont {J.}~\bibnamefont
  {Egge}},\ }\bibfield  {title} {\bibinfo {title} {Axion haloscope signal power
  from reciprocity},\ }\href {https://doi.org/10.1088/1475-7516/2023/04/064}
  {\bibfield  {journal} {\bibinfo  {journal} {Journal of Cosmology and
  Astroparticle Physics}\ }\textbf {\bibinfo {volume} {2023}}\bibinfo  {number}
  { (04)},\ \bibinfo {pages} {064}}\BibitemShut {NoStop}%
\bibitem [{\citenamefont {Egge}\ \emph {et~al.}(2024)\citenamefont {Egge},
  \citenamefont {Ekmedžić}, \citenamefont {Gardikiotis}, \citenamefont
  {Garutti}, \citenamefont {Heyminck}, \citenamefont {Kasemann}, \citenamefont
  {Knirck}, \citenamefont {Kramer}, \citenamefont {Krieger}, \citenamefont
  {Leppla-Weber}, \citenamefont {Martens}, \citenamefont {Öz}, \citenamefont
  {Salama}, \citenamefont {Schmidt}, \citenamefont {Wang},\ and\ \citenamefont
  {Wieching}}]{Egge_2024}%
  \BibitemOpen
\bibfield  {number} {  }\bibfield  {author} {\bibinfo {author} {\bibfnamefont
  {J.}~\bibnamefont {Egge}}, \bibinfo {author} {\bibfnamefont {M.}~\bibnamefont
  {Ekmedžić}}, \bibinfo {author} {\bibfnamefont {A.}~\bibnamefont
  {Gardikiotis}}, \bibinfo {author} {\bibfnamefont {E.}~\bibnamefont
  {Garutti}}, \bibinfo {author} {\bibfnamefont {S.}~\bibnamefont {Heyminck}},
  \bibinfo {author} {\bibfnamefont {C.}~\bibnamefont {Kasemann}}, \bibinfo
  {author} {\bibfnamefont {S.}~\bibnamefont {Knirck}}, \bibinfo {author}
  {\bibfnamefont {M.}~\bibnamefont {Kramer}}, \bibinfo {author} {\bibfnamefont
  {C.}~\bibnamefont {Krieger}}, \bibinfo {author} {\bibfnamefont
  {D.}~\bibnamefont {Leppla-Weber}}, \bibinfo {author} {\bibfnamefont
  {S.}~\bibnamefont {Martens}}, \bibinfo {author} {\bibfnamefont
  {E.}~\bibnamefont {Öz}}, \bibinfo {author} {\bibfnamefont {N.}~\bibnamefont
  {Salama}}, \bibinfo {author} {\bibfnamefont {A.}~\bibnamefont {Schmidt}},
  \bibinfo {author} {\bibfnamefont {H.}~\bibnamefont {Wang}},\ and\ \bibinfo
  {author} {\bibfnamefont {G.}~\bibnamefont {Wieching}},\ }\bibfield  {title}
  {\bibinfo {title} {Experimental determination of axion signal power of dish
  antennas and dielectric haloscopes using the reciprocity approach},\ }\href
  {https://doi.org/10.1088/1475-7516/2024/04/005} {\bibfield  {journal}
  {\bibinfo  {journal} {Journal of Cosmology and Astroparticle Physics}\
  }\textbf {\bibinfo {volume} {2024}}\bibinfo  {number} { (04)},\ \bibinfo
  {pages} {005}}\BibitemShut {NoStop}%
\end{thebibliography}%

\end{document}